\documentclass[onecolumn, draftclsnofoot, 12pt]{IEEEtran}
\pdfoutput=1 
\usepackage{amsmath}
\usepackage{cite}
\usepackage{amssymb}
\usepackage{amsfonts}
\usepackage{algorithm}
\usepackage{algpseudocode}
\usepackage{dsfont}
\usepackage{graphicx}
\usepackage{epsfig}
\usepackage{subfigure}
\usepackage{psfrag}
\usepackage{xcolor}
\usepackage{url}
\usepackage[colorlinks,linkcolor=black,urlcolor=black,anchorcolor=black,citecolor=black,hyperfootnotes=true]{hyperref}
\usepackage{tikz}
\usepackage{stfloats}

\usepackage{enumitem}
\usepackage{setspace}

\newtheorem{theorem}{Theorem}

\newcommand{\mv}[1]{\mbox{\boldmath{$ #1 $}}}

\title{Device Activity Detection in mMTC with Low-Resolution ADC: A New Protocol}
\author{Zhaorui Wang, Ya-Feng Liu, Ziyue Wang, Liang Liu, Haoyuan Pan, and Shuguang Cui
\thanks{Z. Wang and S. Cui are with the Future Network of Intelligence Institute, and the School of Science and Engineering, The Chinese University of Hong Kong, Shenzhen, China, 518172 (e-mail: \{wangzhaorui, shuguangcui\}@cuhk.edu.cn). Y.-F. Liu and Z. Wang are with the State Key Laboratory of Scientific and Engineering Computing, Institute of Computational Mathematics and Scientific/Engineering Computing, Academy of Mathematics and Systems Science, Chinese Academy of Sciences, Beijing 100190, China (e-mail: \{yafliu, ziyuewang\}@lsec.cc.ac.cn). L. Liu is with the Department of Electronic and Information Engineering, The Hong Kong Polytechnic University, Hong Kong (e-mail: liang-eie.liu@polyu.edu.hk). H. Pan is with College of Computer Science and Software Engineering, Shenzhen University, Shenzhen, China (e-mail: hypan@szu.edu.cn).}}

\begin{document}
\maketitle \thispagestyle{empty}

\begin{abstract}
This paper investigates the effect of low-resolution analog-to-digital converters (ADCs) on device activity detection in massive machine-type communications (mMTC). The low-resolution ADCs induce two  challenges on the device activity detection compared with the traditional setup with the assumption of infinite ADC resolution. First, the codebook design for signal quantization by the low-resolution ADC is particularly important since a good design of the codebook can lead to small quantization error on the received signal, which in turn has significant influence on the activity detector performance. To this end, prior information about the received signal power is needed, which depends on the number of active devices $K$. This is sharply different from the activity detection problem in traditional setups, in which the knowledge of $K$ is not required by the BS as a prerequisite. Second, the covariance-based approach achieves good activity detection performance in traditional setups while it is not clear if it can still achieve good performance in this paper. To solve the above challenges, we propose a communication protocol that consists of an estimator for $K$ and a detector for active device identities: 1) For the estimator, the technical difficulty is that the design of the ADC quantizer and the estimation of $K$ are closely intertwined and doing one needs the information/execution from the other. We propose a progressive estimator which iteratively performs the estimation of  $K$ and the design of the ADC quantizer; 2) For the activity detector, we propose a custom-designed stochastic gradient descent algorithm to estimate the active device identities. Numerical results demonstrate the effectiveness of the communication protocol. 
\end{abstract}


\section{Introduction}\label{sec:Introduction}

\subsection{Motivation}
Massive machine-type communications (mMTC) provide efficient random access for a large number of devices, out of which only a small number of them are active\cite{bockelmann2016massive,chen2020massive}. For example, a lot of sensors are deployed in a metering system, while only a small group of them is active to report their status to a BS. mMTC is expected to play an essential role in 5G/6G systems, with the promise to revolutionize manufacturing, healthcare, logistics, and process automation in our future world\cite{shariatmadari2015machine}. A key challenge in mMTC is how to efficiently identify the active devices since the knowledge of the device activity is quite important for the subsequent signal processing.

There are generally two kinds of approaches to detect active devices. One is the compressed sensing (CS) technique by taking advantage of the sporadic nature of the device activity pattern \cite{senel2018grant,chen2018sparse,jiang2018joint,ke2020compressive,ding2019sparsity,sun2019exploiting,mei2021compressive,liu2018massive}. The other is the covariance-based method by making use of the covariance of received signals\cite{Chenphase,haghighatshoar2018improved,chen2021sparse,chen2019covariance,wang2021efficient,ganesan2020algorithm,accwang,wang2022covariance,li2022asynchronous,wang2022scaling}. First, \cite{liu2018massive,Chenphase} have shown that, the massive multiple-input and multiple-output (MIMO) technique can improve the performance of the device activity detector significantly. Second, for the same activity detection performance,  the minimum preamble sequence length required by the covariance-based approach is much smaller than that by the CS approach\cite{haghighatshoar2018improved,Chenphase}. In this paper, considering the merits of the massive MIMO and covariance-based approach, we study the active device detection problem through the covariance-based method in mMTC enabled by massive MIMO.

Nowadays, the massive MIMO has been widely deployed in 5G cellular systems, and will be a key enabler in 6G cellular systems since the massive MIMO can provide huge beamforming gain\cite{bjornson2017massive}. However, there are several problems in massive MIMO itself. Specifically, in MIMO systems, each antenna equips with a RF chain which consists of an analog-to-digital converter (ADC). The ADC converts analog signals to digital signals for baseband processing. The traditional MIMO system applies high-resolution ADCs for good signal quantization performance. Note that the high-resolution ADC is not only expensive, but also power consuming\cite{li2007system,walden1999analog,le2005analog}.  When it comes to massive MIMO, since the number of antennas is quite large, the number of ADCs is also quite large. In this case, if we still apply high-resolution ADC in the massive MIMO systems, the overall cost would be quite huge.  Thus, to solve the above problems while still harvesting the beamforming gain from massive MIMO, \cite{mo2015capacity,mo2017channel} propose low-resolution ADCs in massive MIMO. 

In this paper, we study the device activity detection problem in massive MIMO under the low-resolution ADCs to reduce the overall cost. This setup makes the treatment for the device activity detection problem different from that in the traditional setups which assume ADCs have infinite-resolution\cite{senel2018grant,liu2018massive,chen2018sparse,jiang2018joint,ke2020compressive,ding2019sparsity,sun2019exploiting,mei2021compressive,chen2019covariance,haghighatshoar2018improved,Chenphase,wang2021efficient,ganesan2020algorithm,chen2021sparse,accwang,wang2022covariance}, for the following two aspects. First, in this paper, the codebook design for received-signal quantization by the low-resolution ADC is particularly important since a good design of the codebook can lead to small quantization error of the received signal, which in turn has significant influence on the activity detector performance. To this end, prior information about the received signal power is needed, which depends on the number of active devices $K$. That is, a large $K$ possibly induces large received signal power, and vice versa. This is sharply different from the activity detection problem in the traditional setup, in which the knowledge of $K$ is not required by the BS as a prerequisite. In this case, how to get the knowledge of $K$ at the  BS  is the first challenge.

Second, the covariance-based approach achieves good activity detection performance in traditional setups since the covariance matrix of the received signals can be well approximated by the  received-quantized  signals under ADC with infinite-resolution in massive MIMO. However, in this paper, since the received signals are quantized under low-resolution ADCs, the approximation error of the covariance matrix would be quite large. In this case, how to design an activity detector based on the received-quantized signals under low-resolution ADC  is the second challenge. 

\subsection{Contributions}
In this paper, we consider the activity detection problem in mMTC with low-resolution ADC and propose efficient approaches to solving the corresponding detection problem, which thus address the above two challenges. The main contributions of this paper are summarized as follows. 

First, we put forth a communication protocol to solve the detection problem where the antennas at the BS have low-resolution ADCs. The overall communication is divided into two phases. In phase I, we design an estimator for the number of  active devices $K$. Based on the estimated $K$, the BS designs the ADC quantizer accordingly; In phase II, the BS  detects the active device identities based on the received signals quantized by low-resolution ADCs.

Second, we design a detector for active devices in Phase II of the communication protocol,  assuming that the BS already has an estimation of $K$. In the traditional setup with infinite-resolution ADC, the activity detection problem can be efficiently solved by the traditional gradient descent algorithm. However, the gradient in this paper is hard to compute due to the low-resolution ADC sampling. We solve this problem through a custom-designed stochastic gradient descent (SGD) algorithm\cite{bottou2018optimization} by  exploiting the structure of the problem judiciously. We evaluate the performance of the designed detector. We test the detector under correlated Rayleigh fading channels. The numerical results show that, when $K$ is perfectly estimated by the BS, by only applying a 4-bit ADC quantizer, our detector can achieve nearly the same performance as that with infinite-resolution ADC quantizer. We also evaluate the performance of the detector when $K$ is imperfectly estimated by  the BS. The numerical results show that the imperfect knowledge of $K$ can have a significant penalty on the performance of the detector.

Third, we propose an estimator for $K$ in Phase I of the communication protocol. In Phase I, the received signals to estimate $K$ still need to go through the ADC quantizer, while the design of which still need the information of $K$. Thus, the technical difficulty here is that the design of the ADC quantizer and the estimation of $K$ are closely intertwined and doing one needs the information/execution from the other.  We propose a progressive estimator, which iteratively performs the estimation of the number of the active devices $K$ and the design of the quantizer. The numerical results demonstrate the effectiveness of the proposed estimator. 

\subsection{Related Work}

For the device activity detection problem under infinite-resolution ADC, there are generally two lines of research. The first line of research is to  apply the compressed sensing (CS) technique by taking advantage of the sporadic nature of the device activity pattern \cite{senel2018grant,liu2018massive,chen2018sparse,jiang2018joint,ke2020compressive,ding2019sparsity,sun2019exploiting,mei2021compressive}. The second line of research is to apply the covariance-based approach\cite{haghighatshoar2018improved,chen2019covariance,Chenphase,wang2021efficient,ganesan2020algorithm,chen2021sparse,accwang,wang2022covariance,li2022asynchronous,wang2022scaling}.
Specifically,  the activity can be detected merely based on the covariance matrix of the received  signals since the active device pattern is contained in the covariance matrix. Unlike \cite{senel2018grant,liu2018massive,chen2018sparse,jiang2018joint,ke2020compressive,ding2019sparsity,sun2019exploiting,mei2021compressive,haghighatshoar2018improved,chen2019covariance,Chenphase,wang2021efficient,ganesan2020algorithm,chen2021sparse,accwang,wang2022covariance,li2022asynchronous,wang2022scaling}, we study the device activity detection problem under  low-resolution ADCs at the BS.

For the device activity detection problem under low-resolution ADC, there are few works\cite{yang2019joint,liu2019generalized,mei2022massive}. Specifically, \cite{yang2019joint} studied the problem when the BS has low-resolution ADCs  and the end devices have 1-bit ADCs. In addition, \cite{liu2019generalized,mei2022massive} studied the problem under mixed ADCs at the BS, in which parts of ADCs have infinite-resolution and the others have low-resolution. The key differences between this paper and the works in \cite{yang2019joint,liu2019generalized,mei2022massive} are twofold. First, \cite{yang2019joint,liu2019generalized,mei2022massive} solved the problem by applying the CS approach. Unlike them, we design a detector for active devices through the covariance-based approach. Second, \cite{yang2019joint,liu2019generalized,mei2022massive} implicitly assume that the knowledge of the number of active devices $K$ is perfectly known at the BS as a priori.  Unlike them, we  consider a more general case where  the knowledge of $K$ and its statistics are unknown at the BS. We demonstrate the significant effect of the knowledge $K$ on the activity detection performance, and  propose an estimator to estimate $K$. 

\subsection{Organization}
The rest of this paper is organized as follows. In Section \ref{sec:SYS}, we introduce the system model, and show that the ADC quantizer design is affected by  $K$. In addition, we also propose a communication protocol that consists of an estimator for $K$ and an detector for device identities. Assuming the knowledge of $K$ has been estimated by the BS, we design a detector for active devices in Section \ref{sec:SGD}.  We also evaluate the performance of the designed detector in Section \ref{sec:num1}. In Section \ref{sec: es_num},
we design an estimator for  $K$. We also evaluate the performance of the estimator and the overall communication protocol. Finally, we conclude this paper in Section \ref{sec:col}.


\section{System Model and Communication Protocol}\label{sec:SYS}
\subsection{System Model}
We study a narrowband uplink  system which consists of a BS equipped with $M$ antennas and $N$ single-antenna devices. The propagation distance is assumed to be far larger than the array aperture at the BS so that the devices are in the far-field of the array.  Due to the sporadic traffic,  only $K\ll N$ devices are active at each coherence block. Let $\alpha_n$ be an activity indicator of device $n$, and 
\begin{equation}
	\alpha_n=\left\{\begin{array} {ll} 1, ~ {\rm if ~ device} ~ n ~ {\rm is ~ active}, \\ 0, ~ {\rm otherwise}, \end{array} ~~n=1, 2, \dots, N.\right. 
\end{equation}
We assume block-fading channels, i.e., the channels remain roughly constant within each coherence block, but may vary among different coherence blocks. Let $\sqrt{\beta_n}\mv{h}_n\in\mathbb{C}^{M\times 1}$ be the channel between device $n$ and the BS, where $\beta_n$ is the large-scale fading component due to path-loss and shadowing. In this paper, $\mv{h}_n$ is modeled as correlated Rayleigh fading channel,  i.e.,
\begin{align}
	\mv{h}_n=\mv{C}_n^{\frac{1}{2}}\ddot{\mv{h}}_n, ~n=1, 2, \dots,N,\label{eq:cor}
\end{align}
where $\mv{C}_n$  denotes the channel covariance matrix of $\mv{h}_n$ and diagonal elements of $\mv{C}_n$ are all ones \cite{bjornson2017massive}, and $\ddot{\mv{h}}_n\sim\mathcal{CN}(0,\mv{I})$  follows the independent and identically distributed (i.i.d.) Rayleigh fading channel model. In addition, we assume that the channels among different devices are independent with each other, i.e., $\mathbb{E}\left(\sqrt{\beta_n}\mv{h}_n\sqrt{\beta_{n'}}\mv{h}^H_{n'}\right)=\mv{0}$, for $n\ne n'$\cite{bjornson2017massive,bjornson2015massive}, where $\mathbb{E}(x)$ denotes the expectation of a random variable $x$.

Each device is preassigned a unique and non-orthogonal preamble sequence $\mv{s}_n\in\mathbb{C}^{L_{\rm I}\times 1}$ for active device identity detection at the BS, $n=1, 2, \dots, N$, where $L_{\rm I}\ll N$ is the preamble sequence length for active device \textbf{I}dentity detection. The sequences $\mv{s}_n$'s are known as a priori at the BS. The received baseband-equivalent signal $\mv{Y}\in\mathbb{C}^{L_{\rm I}\times M}$ from time slot\footnote{Note that one time slot equals one symbol duration in the preamble sequence.} 1 to time slot $L_{\rm I}$ is expressed as
\begin{align}
	\mv{Y}&=\left[\mv{y}_1, \mv{y}_2, \dots,\mv{y}_M\right]\nonumber\\
	&=\sum_{n=1}^{N}\alpha_n\mv{s}_n\sqrt{p_n}\sqrt{\beta_n}\mv{h}^T_n+\mv{Z}\label{eq:a}\\
	&=\sum_{n=1}^{N}\alpha_n\mv{s}_n\sqrt{\beta}\mv{h}^T_n+\mv{Z}\label{eq:b}\\
	&=\mv{S}\mv{\gamma}^{\frac{1}{2}}\mv{H}+\mv{Z},\label{eq:rec}
\end{align}
where $\mv{y}_m\in\mathbb{C}^{L_{\rm I}\times 1}$ is the received signal from time slot 1 to time slot $L_{\rm I}$ at the $m$-th antenna, $\forall m$; $p_n$ is the transmit power of device $n$; $\mv{Z}=\left[\mv{z}_1, \mv{z}_2, \dots,\mv{z}_M\right]\in\mathbb{C}^{L_{\rm I}\times M}$ is the additive white Gaussian noise with $\mv{z}_m\sim \mathcal{CN}(0,\sigma^2\mv{I})$, $\forall m$. Due to the power control on each device, in \eqref{eq:b} we have
\begin{align}
	\beta=p_n\beta_n,~n=1,\dots, N;\label{eq:beta}
\end{align}
$\mv{S}=\left[\mv{s}_1, \mv{s}_2, \dots,\mv{s}_N\right]\in\mathbb{C}^{L_{\rm I}\times N}$ is the preamble matrix consisting of the preambles from different devices; $\mv{\gamma}={\rm diag}\left\{\alpha_1\beta, \alpha_2\beta, \dots, \alpha_N\beta\right\}\in\mathbb{R}^{N\times N}$; and  $\mv{H}=\left[\mv{h}_1, \mv{h}_2, \dots,\mv{h}_N\right]^T\in\mathbb{C}^{N\times M}$ is the channel matrix. In this paper, the preamble matrix $\mv{S}$ is designed as follows. Denote $s_{\ell,n}$  the element in the $\ell$-th row and the $n$-th column of $\mv{S}$, $\Re{(s_{\ell,n})}$ the real part of  $s_{\ell,n}$, and  $\Im{(s_{\ell,n})}$  the imaginary part of  $s_{\ell,n}$. Then, both $\Re{(s_{\ell,n})}$ and $\Im{(s_{\ell,n})}$ are uniformly chosen from  $\{-\frac{\sqrt{2}}{2},\frac{\sqrt{2}}{2}\}$, $\forall n, \ell$. The preamble matrix $\mv{S}$ is assumed to be known at the BS.

In the following, we rewrite the received signal $\mv{Y}$ in \eqref{eq:rec} in order to show the ADC quantizer design. In \eqref{eq:rec}, denote $\hat{\mv{y}}_m=\left[\Re{(\mv{y}^T_m)},  \Im{(\mv{y}^T_m)}\right]^T\in\mathbb{R}^{2L_{\rm I}\times 1}$ and  $\bar{\mv{y}}=\left[\hat{\mv{y}}_1^T, \hat{\mv{y}}_2^T, \dots, \hat{\mv{y}}_M^T\right]^T\in\mathbb{R}^{2L_{\rm I}M\times 1}$. In addition, denote $\mv{h}^{\rm col}_m$  the $m$-th column of $\mv{H}$ in \eqref{eq:rec}, $\hat{\mv{h}}_m=\left[\Re{(\mv{h}^{\rm col}_m)^T}, \Im{(\mv{h}^{\rm col}_m)^T}\right]^T\in\mathbb{R}^{2N\times 1}$, and $\bar{\mv{h}}=\left[\hat{\mv{h}}_1^T, \hat{\mv{h}}_2^T, \dots, \hat{\mv{h}}_M^T\right]^T\in\mathbb{R}^{2NM\times 1}$. We have the following Theorem \ref{theorem0}. 

\begin{theorem}\label{theorem0}
	The received signal $\mv{Y}$ in \eqref{eq:rec} is equivalent to 
	\begin{align}
		\bar{\mv{y}}=\bar{\mv{S}}\bar{\mv{\gamma}}^{\frac{1}{2}}\bar{\mv{h}}+\bar{\mv{z}},\label{eq:rec_r}
	\end{align}
	where $\bar{\mv{S}}={\rm diag}\{\underbrace{\hat{\mv{S}}, \dots,\hat{\mv{S}}}_M\}\in\mathbb{R}^{2L_{\rm I}M\times 2NM}$, which consists of $M$ copies of $\hat{\mv{S}}$ in its main-diagonal blocks, and 
	\begin{align}
		\hat{\mv{S}}=\begin{bmatrix}
			\Re{\left(\mv{S}\right)} & -\Im{\left(\mv{S}\right)}\\
			\Im{\left(\mv{S}\right)} & \Re{\left(\mv{S}\right)}
		\end{bmatrix}\in\mathbb{R}^{2L_{\rm I}\times 2N};\label{eq:s_hat}
	\end{align}
	$\bar{\mv{\gamma}}={\rm diag}\{\underbrace{\hat{\mv{\gamma}},\dots,\hat{\mv{\gamma}}}_M\}\in\mathbb{R}^{2NM\times 2NM}$ with $\hat{\mv{\gamma}}={\rm diag}\left\{\mv{\gamma}, \mv{\gamma}\right\}$;  and $\bar{\mv{z}}=\left[\hat{\mv{z}}_1^T, \hat{\mv{z}}_2^T, \dots, \hat{\mv{z}}_M^T\right]^T\in\mathbb{R}^{2L_{\rm I}M\times 1}$ with $\hat{\mv{z}}_m=\left[\Re{(\mv{z}^T_m)}, \Im{(\mv{z}^T_m)}\right]^T\in\mathbb{R}^{2L_{\rm I}\times 1}$. 
\end{theorem}
\begin{IEEEproof}
	Please refer to Appendix \ref{appendix0}.
\end{IEEEproof}

The covariance matrix of $\bar{\mv{y}}$ in \eqref{eq:rec_r} is expressed as
\begin{align}
	\mv{\Sigma}=\mathbb{E}(\mv{\bar{y}}\mv{\bar{y}}^T)=\bar{\mv{S}}\bar{\mv{\gamma}}^{\frac{1}{2}}\mv{C}\bar{\mv{\gamma}}^{\frac{1}{2}}\bar{\mv{S}}^T+\frac{1}{2}\sigma^2\mv{I}=\bar{\mv{S}}\mv{C}\bar{\mv{\gamma}}\bar{\mv{S}}^T+\frac{1}{2}\sigma^2\mv{I}, \label{eq:Sigma}
\end{align}
where $\mv{C}=\mathbb{E}(\mv{\bar{h}}\mv{\bar{h}}^T)$. The derivation of  $\mv{\Sigma}$ in \eqref{eq:Sigma} and the way of constructing $\mv{C}$ from $\mv{C}_n$'s in (2) are  shown in  Appendix \ref{appendix1}.

\begin{theorem}\label{theorem1}
	Let $\lambda^{\ell, \ell}$ denote the power of $\bar y_{\ell}$,~$\ell=1,2,\dots, 2L_{\rm I}M$, where  $\bar{y}_{\ell}$ is the $\ell$-th element of the received signal  $\bar{\mv{y}}$ in \eqref{eq:rec_r}. Then we have
	\begin{align}
		\lambda^{\ell, \ell}=\mathbb{E}\left(\bar y_{\ell}\bar y_{\ell}\right)=\frac{K}{2}\beta+\frac{1}{2}\sigma^2,~~\forall \ell, \label{eq:lambda21}
	\end{align}
where $\beta$ has been defined in \eqref{eq:beta}. 
\end{theorem}
\begin{IEEEproof}
	Please refer to Appendix \ref{appendix2}.
\end{IEEEproof}

Theorem \ref{theorem1} shows that the signal power of each dimension of $\bar{\mv{y}}$ is the same. In this case, in \eqref{eq:lambda21}, denote 
\begin{align}
	\lambda=\lambda^{\ell, \ell}, ~\ell=1,2,\dots, 2L_{\rm I}M.\label{eq: lambda}
\end{align}

\subsection{ADC Quantizer Design}\label{sec: sys_A}
At the BS, the received  signal $\bar{\mv{y}}$ in \eqref{eq:rec_r} is  quantized through ADCs for further baseband processing. Specifically, within each ADC, denote $Q(\bar{y}_\ell)$ a $B$-bit quantizer of $\bar{y}_\ell$, $\ell=1,2, \dots, 2L_{\rm I}M$, where $B$ specifies the resolution of the quantizer. In this paper, we apply a uniform quantizer $Q(\bar{y}_\ell)$ in ADC like that in \cite{yang2019joint,liu2019generalized,mei2022massive}. Note that the uniform ADC quantizers have been widely applied within the receivers in  current wireless systems. Denote the quantization step size by $\Delta$. Then, the $q$-th quantization interval in $Q(\bar{y}_\ell)$ is
\begin{align}
	\mathcal{I}_q=\left\{\begin{array} {lll} \left(-\infty, (1-2^{(B-1)})\Delta\right), ~~~~~~~~~~~~~~~~~{\rm if} ~ q=1,
	\\ \left[(q-1-2^{(B-1)})\Delta, (q-2^{(B-1)})\Delta\right), ~ {\rm if}~q=2,\dots,2^B-1, 
	\\\left[(2^{(B-1)}-1)\Delta, +\infty\right),~~~~~~~~~~~~~~~~~~{\rm if} ~ q=2^B. \end{array}\right.\label{eq:q_level}
\end{align} 
In this case, if the input signal $\bar{y}_\ell\in\mathcal{I}_q$, $\ell=1,2, \dots, 2L_{\rm I}M$, then the output signal after quantization is
\begin{align}
	Q(\bar{y}_\ell)=\left\{\begin{array} {lll} (\frac{1}{2}-2^{(B-1)})\Delta, ~~~~~~ {\rm if} ~ q=1, \\ (q-2^{(B-1)}-\frac{1}{2})\Delta, ~ {\rm if}~q=2,\dots,2^B-1, \\(2^{(B-1)}-\frac{1}{2})\Delta,~~~~~~ {\rm if} ~ q=2^B.\end{array}\right.\label{eq:codebook}
\end{align}
With the above ADC quantizer, the received-quantized signal is expressed as
\begin{align}
	\bar{\mv{y}}^{Q}=\left[\bar{y}_1^{Q}, \bar{y}_2^{Q},\dots, \bar{y}_{2L_{\rm I}M}^{Q} \right]^T,
\end{align}
where 
\begin{align}
	\bar{y}_\ell^{Q}=Q(\bar{y}_\ell), ~~\ell=1, 2, \dots, 2L_{\rm I}M.\label{eq: y_re}
\end{align}

In the following, we show the design of the quantization step size $\Delta$ in the ADC. From \eqref{eq:rec_r} and \eqref{eq:lambda21},  $\bar{y}_\ell$ is Gaussian distributed with mean zero  and power (i.e., the variance) $\lambda$, $\forall \ell$. In this case, most of the realizations of $\bar{y}_\ell$ are distributed around the mean zero. For example, $99.7\%$ of the realizations of $\bar{y}_\ell$ fall in the interval $[-3\sqrt{\lambda}, 3\sqrt{\lambda}]$. In this case,  the quantization step size should be set such that the received signal $\bar{y}_{\ell}$ that falls in the interval $[-3\sqrt{\lambda}, 3\sqrt{\lambda}]$ can be uniformly quantized. Thus, the step size in the uniform ADC quantizer $Q(\bar{y}_\ell)$ in \eqref{eq:codebook} is expressed as
\begin{align}
	\Delta=\frac{\rho\sqrt{\lambda}}{2^{B-1}}=\frac{\rho\sqrt{2K\beta+2\sigma^2}}{2^B}, \label{eq:step2}
\end{align}
where $\rho>0$ is a parameter that controls the size of the sample space along $\bar{y}_{\ell}$, $\ell=1, 2, \dots, 2L_{\rm I}M$.  From \eqref{eq:step2}, given the quantization bits $B$,  a signal with large power $\lambda$ should be quantized by a  quantizer with large quantization step size $\Delta$, and vice versa. As such, most realizations of the received signals can be well sampled by the uniform ADC quantizer.  Based on the quantization step size $\Delta$ in \eqref{eq:step2}, we can design the codebook according to \eqref{eq:codebook}. 

As shown in \eqref{eq:step2}, the quantization step size $\Delta$ depends on the number of active devices $K$. In particular, a considerably overestimated or underestimated of $K$ by the BS would result in huge ADC quantization error of $\bar{\mv{y}}$, and hence deteriorates the performance of the active device detector. In Section \ref{sec:Kes}, we will numerically show that the imperfect knowledge of $K$ can have a significant penalty on the performance of the detector. Therefore, the accurate knowledge of $K$ is quite important for the detector.. This is quite different from the current studies \cite{senel2018grant,liu2018massive,chen2018sparse,jiang2018joint,ke2020compressive,ding2019sparsity,sun2019exploiting,mei2021compressive,chen2019covariance,haghighatshoar2018improved,Chenphase,wang2021efficient,ganesan2020algorithm,chen2021sparse,accwang,wang2022covariance} in mMTC in which they assume ADC with infinite-resolution and detect the device activities directly without the knowledge of $K$.

\subsection{Communication Protocol}\label{sec:pro}

\begin{figure}[t]
	\centering
	\includegraphics[width=10cm]{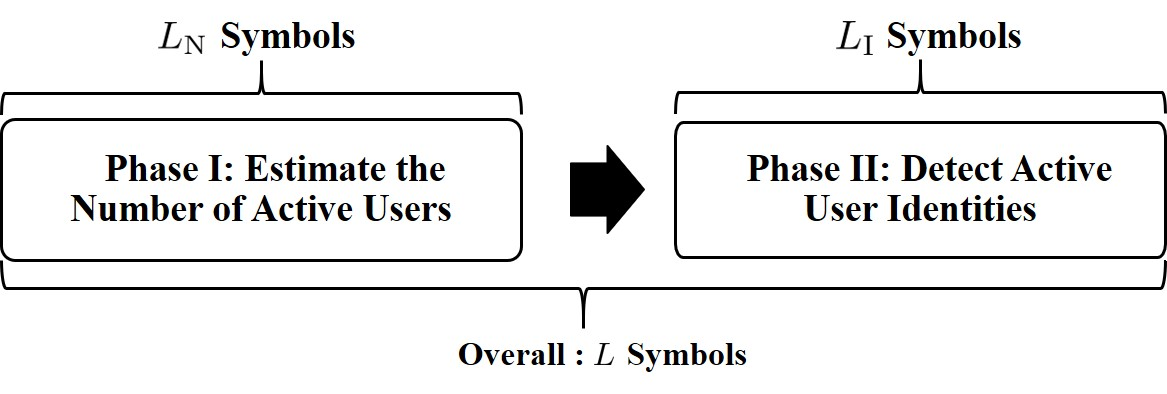}
	\caption{The overall communication protocol.}\label{Figpro}
\end{figure}

To solve this problem, we propose an estimator to estimate $K$ before the device activity detector. As shown in Fig. \ref{Figpro}, the overall communication protocol is divided into the two phases:
\begin{enumerate}
	\item \textbf{Phase I:} The BS spends $L_{\rm N}$ symbols to estimate the \textbf{N}umber of active devices. The estimated number of active devices is denoted by $\hat{K}$. Based on $\hat{K}$, the BS  designs the quantization codebook according to \eqref{eq:step2} and \eqref{eq:codebook} (by replacing $K$ by $\hat{K}$ therein).
	\item \textbf{Phase II:} The devices report their activity through $L_{\rm I}$ preambles. Then, the BS  detects the active devices based on the received signals quantized by the designed ADC quantizer. 
\end{enumerate}
The total time slots used for detection and estimation are
\begin{align}
	L=L_{\rm N}+L_{\rm I}.
\end{align}

In Section \ref{sec:SGD} of this paper, we first show the design of the device activity detector in Phase II  of the communication protocol, assuming the BS has the knowledge of $\hat{K}$ from Phase I. Through the numerical results in Section \ref{sec:num1}, we show that the knowledge of $\hat{K}$ indeed has significant influence on the performance of  the activity detector. However, fortunately, the numerical results also show that we actually do not need an accurate estimate of $K$. Instead,  a rough estimate of $K$ is good enough for the device activity detector in Phase II. Then, building on this important observation, we show the design of the estimator for $K$ in Phase I in Section \ref{sec: es_num}.

\section{Device Activity Detector in Phase II}\label{sec:SGD}

In this section, we study the device activity detector in Phase II of the communication protocol shown in Fig. \ref{Figpro}, under the assumption that the BS has the estimated knowledge $\hat{K}$ from Phase I. In this case, in Phase II, the received signal $\bar{\mv{y}}$ to detect activity is now quantized by the ADC quantizer  according to \eqref{eq:codebook} and \eqref{eq:step2} (by replacing $K$ by $\hat{K}$ therein). We show how to estimate $\mv{\gamma}$ in \eqref{eq:rec} based on the received-quantized signal $\bar{\mv{y}}^{Q}$ in \eqref{eq: y_re}. Once $\mv{\gamma}$ is estimated, we can determine the active devices based on the estimated $\mv{\gamma}$. 

\subsection{Problem Formulation}
Denote $p(\bar{\mv{y}}^{Q}
|\mv{\gamma},\mv{S})$ the probability density function (PDF) of $\bar{\mv{y}}^{Q}$ given $\mv{\gamma}$ and $\mv{S}$. The detection problem is formulated as
\begin{alignat}{3}
	&\min_{\mv{\gamma}} &\quad &-p(\bar{\mv{y}}^{Q}
	\mid\mv{\gamma},\mv{S}) \label{eq:P2}\\
	&~~\text{s.t.}                          &      & \mv{\gamma}\ge 0.\nonumber
\end{alignat}
The PDF $p(\mv{y}^{Q}
\mid\mv{\gamma},\mv{S})$ in \eqref{eq:P2} is expressed as
\begin{align}
	p(\bar{\mv{y}}^{Q}\mid\mv{\gamma},\mv{S})=\int_{\mathcal{I}_{f(\bar{y}_{1}^{Q})}}\int_{\mathcal{I}_{f(\bar{y}_{2}^{Q})}}\dots\int_{\mathcal{I}_{f(\bar{y}_{2L_{\rm I}M}^{Q})}}p(\bar{\mv{y}}\mid\mv{\gamma},\mv{S})d\bar{y}_{1}d\bar{y}_{2}\dots d\bar{y}_{2L_{\rm I}M},\label{eq:ml}
\end{align}
where $f(x)$ maps $x$ to the index of the quantization interval where it falls in, i.e.,
\begin{align}
	f(x)=\left\{\begin{array} {lll} 1, ~~~~{\rm if} ~ x\in\left(-\infty, (1-2^{(B-1)})\Delta\right), \\ q, ~~~~{\rm if}~x\in\left[(q-1-2^{(B-1)})\Delta, (q-2^{(B-1)})\Delta\right), ~q\in[2, 2^B-1], \\2^B,~~{\rm if} ~x\in\left[(2^{(B-1)}-1)\Delta, +\infty\right), \end{array}\right.\label{eq:f}
\end{align}
and  $p(\bar{\mv{y}}\mid\mv{\gamma},\mv{S})$ denotes the PDF of $\bar{\mv{y}}$ given $\mv{\gamma}$ and  $\mv{S}$, i.e.,
\begin{align}
	p(\bar{\mv{y}}\mid\mv{\gamma},\mv{S})=\frac{1}{(2\pi)^{ML}|\mv{\Sigma}|^{\frac{1}{2}}}\exp{\left(-\frac{1}{2}\bar{\mv{y}}^T\mv{\Sigma}^{-1}\bar{\mv{y}}\right)}.\label{eq:pdf1}
\end{align}
The expression of $\mv{\Sigma}$ is shown in \eqref{eq:Sigma}.

\subsection{Computational Challenge of Problem \eqref{eq:P2} and Motivation of SGD Algorithm}

In the rest of this section, we show how to solve the estimation problem in \eqref{eq:P2}. Specifically, if the ADC resolution is infinite, then the objective function $p(\bar{\mv{y}}^{Q}\mid\mv{\gamma},\mv{S})$ in \eqref{eq:P2} reduces to
\begin{align}
	p(\bar{\mv{y}}^{Q}\mid\mv{\gamma},\mv{S})=p(\bar{\mv{y}}\mid\mv{\gamma},\mv{S}).\label{eq:m4}
\end{align}
In this case, the problem in \eqref{eq:P2} could be solved efficiently by the traditional gradient descent algorithm\cite{birgin2000nonmonotone,wang2021efficient}, where the gradient of 	$p(\bar{\mv{y}}^{Q}|\mv{\gamma},\mv{S})$ with respect to $\mv{\gamma}$ can be computed with ${\cal O}(NL_{\rm I}^2)$ complexity.  However, in this paper when the ADC resolution is finite, the objective function $p(\bar{\mv{y}}^{Q}\mid\mv{\gamma},\mv{S})$ is an integral of $p(\bar{\mv{y}}\mid\mv{\gamma},\mv{S})$ over the space
\begin{align}
\mathcal{I}\left(\bar{y}_{1}^{Q}, \bar{y}_{2}^{Q}, \dots,\bar{y}_{2L_{\rm I}M}^{Q}\right)\overset{\Delta}{=}\mathcal{I}_{f(\bar{y}_{1}^{Q})}\times\mathcal{I}_{f(\bar{y}_{2}^{Q})}\times\dots\times\mathcal{I}_{f(\bar{y}_{2L_{\rm I}M}^{Q})},\label{eq:sp}
\end{align}
and its gradient is also an integral of the gradient of $p(\bar{\mv{y}}\mid\mv{\gamma},\mv{S})$ over the space $\mathcal{I}\left(\bar{y}_{1}^{Q},\dots,\bar{y}_{2L_{\rm I}M}^{Q}\right)$, i.e., 
\begin{align}
	\nabla p(\bar{\mv{y}}^{Q}\mid\mv{\gamma},\mv{S})=\int_{\mathcal{I}_{f(\bar{y}_{1}^{Q})}}\int_{\mathcal{I}_{f(\bar{y}_{2}^{Q})}}\dots\int_{\mathcal{I}_{f(\bar{y}_{2L_{\rm I}M}^{Q})}}\nabla p(\bar{\mv{y}}\mid\mv{\gamma},\mv{S})d\bar{y}_{1}d\bar{y}_{2}\dots d\bar{y}_{2L_{\rm I}M},\label{eq:m5}
\end{align}
where $\nabla p(\mv{x})$ denotes the gradient of $p(\mv{x})$. 

It is challenging to compute the gradient in \eqref{eq:m5} and apply the traditional gradient descent algorithm to solve problem \eqref{eq:P2}. More specifically, although we can get a closed-form of $\nabla p(\bar{\mv{y}}|\mv{\gamma},\mv{S})$ in \eqref{eq:m5}, it is hard to get a closed-form of $\nabla p(\bar{\mv{y}}^{Q}\mid\mv{\gamma},\mv{S})$ due to the high-dimensional integration (especially in the massive MIMO system where $M$ is usually large). As an alternative, one can numerically compute the gradient in \eqref{eq:m5}. However, the computational complexity is quite high, which further prevents us from applying the traditional gradient descent algorithm to solve the problem \eqref{eq:P2}. Specifically, if we sample $\bar{\mv{y}}$ uniformly, the gradient $\nabla p(\bar{\mv{y}}^{Q}|\mv{\gamma},\mv{S})$ in \eqref{eq:m5} can be approximated as 
\begin{align}
	\nabla p(\bar{\mv{y}}^{Q}\mid\mv{\gamma},\mv{S})\approx\mu\sum_{i_1}\sum_{i_2}\dots\sum_{i_{2L_{\rm I}M}}\nabla p\left(\left[\bar{y}_1^{(i_1)}, \bar{y}_2^{(i_2)}, \dots, \bar{y}_{2L_{\rm I}M}^{(i_{2L_{\rm I}M})}\right]^T\Big|\mv{\gamma},\mv{S}\right),\label{eq:m6}
\end{align} 
where $\mu$ is a constant, and $\bar{y}_{\ell}^{(i_{\ell})}$ is the $i_{\ell}$-th sample of $\bar{y}_{\ell}$ in the sample space $\mathcal{I}_{f(\bar{y}_{{\ell}}^{Q})}$. Suppose  the number of samples in each dimension is $I$. Then from  \eqref{eq:m6}, we know that the computational complexity is in the order of ${\cal O}(I^{2L_{\rm I}M})$, which would be prohibitively high when $M$ and $L_{\rm I}$ are large. 

To solve the aforementioned computational challenge, we propose to apply a stochastic gradient descent (SGD) algorithm\cite{bottou2018optimization} to solve the problem \eqref{eq:P2}. Specifically, at each iteration, the SGD algorithm computes a stochastic gradient of $p(\bar{\mv{y}}\mid\mv{\gamma},\mv{S})$ by sampling only one point in each dimension of $\bar{\mv{y}}$, instead of sampling many points in each dimension of $\bar{\mv{y}}$ to approximately compute the true gradient like that in \eqref{eq:m6}. The per-iteration computational complexity is thus reduced significantly. Moreover, \cite{bottou2018optimization} has proven the convergence of the SGD algorithm.

In the next two subsections, we tailor the SGD algorithm to solve our interested problem \eqref{eq:P2}, which essentially requires two basic steps. The first step is to rewrite the problem \eqref{eq:P2} into a favorable expectation form required by SGD. The second step is to compute the stochastic gradient, which is still a nontrivial task. 

\subsection{Rewriting \eqref{eq:P2} as a Form of Expected-Value Objective Function}\label{sec:approx}

The SGD algorithm is well suited for solving problems with an expected-value objective function\cite{bottou2018optimization}, e.g., in this paper the objective function should have a form like
\begin{align}
	\mathbb{E}_{\mv{x}}\left(g\left(\mv{x}, \mv{\gamma}\right)\right),\label{eq:ex}
\end{align}
where $g\left(\mv{x}, \mv{\gamma}\right)$ is a function consisting of a random vector $\mv{x}$ and a deterministic vector $\mv{\gamma}$ that needs to be estimated in this paper. In this case, we should rewrite our objective function in \eqref{eq:P2} as an expected-value objective function like that in \eqref{eq:ex}.

First, from \eqref{eq:rec_r} we know that each dimension of $\bar{\mv{y}}$ is Gaussian distributed with zero mean. In particular, for $\bar{y}_{\ell}$, $\ell=1,\dots,2NL_{\rm I}$, although the range of its PDF extends to infinity on both sides, $95\%$ of the realizations of $\bar{y}_{\ell}$ fall in the interval $[-2\sqrt{\lambda}, 2\sqrt{\lambda}]$, and $99.7\%$ of the realizations fall in the interval $[-3\sqrt{\lambda}, 3\sqrt{\lambda}]$, where $\sqrt{\lambda}$ denotes the standard deviation of $\bar{y}_{\ell}$ that has been defined in \eqref{eq: lambda}. In this case, the  objective function $p(\bar{\mv{y}}^{Q}\mid\mv{\gamma},\mv{S})$ in \eqref{eq:P2} can be well approximated if the integration in \eqref{eq:ml} does not take account the values of $\bar{\mv{y}}$ that are far from its mean $\mv{0}$, i.e., 
\begin{align}
	p(\bar{\mv{y}}^{Q}\mid\mv{\gamma},\mv{S})&\approx\int_{\mathcal{J}_{f(\bar{y}_{1}^{Q})}}\int_{\mathcal{J}_{f(\bar{y}_{2}^{Q})}}\dots\int_{\mathcal{J}_{f(\bar{y}_{2L_{\rm I}M}^{Q})}}p(\bar{\mv{y}}\mid\mv{\gamma},\mv{S})d\bar{y}_{1}d\bar{y}_{2}\dots d\bar{y}_{2L_{\rm I}M}\nonumber\\&\overset{\Delta}{=}v\left(\bar{\mv{y}}^{Q}\mid\mv{\gamma},\mv{S}\right),\label{eq:m7}
\end{align}
where 
\begin{align}
	\mathcal{J}_q=\left\{\begin{array} {lll} \left(-2^{(B-1)}\Delta, (1-2^{(B-1)})\Delta\right), ~{\rm if} ~ q=1, \\\left[(2^{(B-1)}-1)\Delta, 2^{(B-1)}\Delta\right),~~~{\rm if} ~ q=2^B, \\ \mathcal{I}_q, ~~~~~~~~~~~~~~~~~~~~~~~~~~~~~~~ {\rm if}~q=2,\dots,2^B-1,\end{array}\right. \label{eq: quan}
\end{align} 
and $f(x)$ has been defined in \eqref{eq:f}.

Second, in \eqref{eq:m7} the equality still holds when replacing the variable $\bar{\mv{y}}$ by another variable  $\mv{x}=[x_1,\dots, x_{2L_{\rm I}M}]^T$ inside the integral, i.e., 
\begin{align}
	v\left(\bar{\mv{y}}^{Q}\mid\mv{\gamma},\mv{S}\right)&=\int_{\mathcal{J}_{f(\bar{y}_{1}^{Q})}}\int_{\mathcal{J}_{f(\bar{y}_{2}^{Q})}}\dots\int_{\mathcal{J}_{f(\bar{y}_{2L_{\rm I}M}^{Q})}}p(\mv{x}\mid\mv{\gamma},\mv{S})dx_{1}dx_{2}\dots dx_{2L_{\rm I}M}\nonumber\\&=\int_{\mathcal{J}_{f(\bar{y}_{1}^{Q})}}\int_{\mathcal{J}_{f(\bar{y}_{2}^{Q})}}\dots\int_{\mathcal{J}_{f(\bar{y}_{2L_{\rm I}M}^{Q})}}p(\mv{x}\mid\mv{\gamma},\mv{S})\frac{1}{p(\mv{x})}p(\mv{x})dx_{1}dx_{2}\dots dx_{2L_{\rm I}M},\label{eq: m31}
\end{align} 
where $p(\mv{x})$ in \eqref{eq: m31} is defined as the PDF of $\mv{x}$. Last, in \eqref{eq: m31} denote
\begin{align}
	g\left(\mv{x}, \mv{\gamma}\right)=p(\mv{x}\mid\mv{\gamma},\mv{S})\frac{1}{p(\mv{x})}. \label{eq:g}
\end{align}
Then, $v\left(\bar{\mv{y}}^{Q}\mid\mv{\gamma},\mv{S}\right)$ in \eqref{eq: m31} is expressed as
\begin{align}
	v\left(\bar{\mv{y}}^{Q}\mid\mv{\gamma},\mv{S}\right)
	&=\int_{\mathcal{J}_{f(\bar{y}_{1}^{Q})}}\int_{\mathcal{J}_{f(\bar{y}_{2}^{Q})}}\dots\int_{\mathcal{J}_{f(\bar{y}_{2L_{\rm I}M}^{Q})}}g\left(\mv{x}, \mv{\gamma}\right)p(\mv{x})dx_{1}dx_{2}\dots dx_{2L_{\rm I}M}\label{eq: m32}\nonumber\\&=\mathbb{E}_{\mv{x}}\left(g\left(\mv{x}, \mv{\gamma}\right)\right).
\end{align}
In particular, in this paper, to ease the algorithm execution and analysis, we define the PDF of $\mv{x}$ as
\begin{align}
p(\mv{x})=\frac{1}{|\mathcal{J}_{f(\bar{y}_{1}^{Q})}||\mathcal{J}_{f(\bar{y}_{2}^{Q})}|\dots|\mathcal{J}_{f(\bar{y}_{2L_{\rm I}M}^{Q})}|}, \label{eq: pdf}
\end{align}
i.e.,  $\mv{x}$ is uniformly distributed in the space
\begin{align}
\mathcal{J}\left(\bar{y}_{1}^{Q},\dots,\bar{y}_{2L_{\rm I}M}^{Q}\right)\overset{\Delta}{=}\mathcal{J}_{f(\bar{y}_{1}^{Q})}\times\mathcal{J}_{f(\bar{y}_{2}^{Q})}\times\dots\times\mathcal{J}_{f(\bar{y}_{2L_{\rm I}M}^{Q})} \label{eq: space}.
\end{align}
The optimization problem in \eqref{eq:P2} is then reformulated as
\begin{alignat}{3}
	&\min_{\mv{\gamma}} &\quad &-\mathbb{E}_{\mv{x}}\left(g\left(\mv{x}, \mv{\gamma}\right)\right) \label{eq:opt2}\\
	&~~\text{s.t.}                          &      & \mv{\gamma}\ge 0,\nonumber
\end{alignat}
where $\mv{x}\in\mathcal{J}\left(\bar{y}_{1}^{Q},\dots,\bar{y}_{2L_{\rm I}M}^{Q}\right)$, and $g\left(\mv{x}, \mv{\gamma}\right)$ is expressed in $\eqref{eq:g}$.

\subsection{Tailored Stochastic Gradient Descent  Algorithm}\label{sec:method}
The SGD algorithm to solve problem \eqref{eq:opt2} is carried out as follows. At the $i$-th iteration, the algorithm uniformly generates a vector $\mv{x}^{(i)}$ from the sample space $\mathcal{J}\left(\bar{y}_{1}^{Q}, \bar{y}_{2}^{Q}, \dots, \bar{y}_{2L_{\rm I}M}^{Q}\right)$  according to its PDF defined in \eqref{eq: pdf}. Then the variable $\mv{\gamma}$ is updated as follows:
\begin{align}
	\mv{\gamma}^{(i)}=\mv{\gamma}^{(i-1)}-\theta_i\nabla g\left(\mv{x}^{(i)}, \mv{\gamma}^{(i-1)}\right), \label{eq: upd}
\end{align}
where $\mv{\gamma}^{(i)}$ is the value of $\mv{\gamma}$ at the $i$-th iteration, $\theta_i$ is a step size parameter that controls the convergence speed of the SGD algorithm, and $\nabla g\left(\mv{x}^{(i)}, \mv{\gamma}^{(i-1)}\right)$ is the stochastic gradient at the $i$-th iteration. The algorithm is terminated when 
\begin{align}
	\frac{||\mv{\gamma}^{(i)}-\mv{\gamma}^{(i-1)}||_1}{N}<\epsilon\label{eq: contol}
\end{align}
where $\epsilon>0$ is a threshold that controls the termination of the algorithm. 

In this paper, it is nontrivial to compute the stochastic gradient $\nabla g\left(\mv{x}^{(i)}, \mv{\gamma}^{(i-1)}\right)$ in \eqref{eq: upd} in order to  execute the SGD algorithm. Specifically,  $\nabla g\left(\mv{x}^{(i)}, \mv{\gamma}^{(i-1)}\right)$ at the $i$-th iteration is expressed as
\begin{align}
	\nabla g\left(\mv{x}^{(i)}, \mv{\gamma}^{(i-1)}\right)=\exp{\left(\bar{g}\left(\mv{x}^{(i)}, \mv{\gamma}^{(i-1)}\right)\right)}\nabla\bar{g}\left(\mv{x}^{(i)}, \mv{\gamma}^{(i-1)}\right), \label{eq:grade}
\end{align}
where
\begin{align}
	\bar{g}\left(\mv{x}^{(i)}, \mv{\gamma}^{(i-1)}\right)&=\log{\left( g\left(\mv{x}^{(i)}, \mv{\gamma}^{(i-1)}\right)\right)}\nonumber\\
	&=-\frac{1}{2}\left(\mv{x}^{(i)}\right)^T\mv{\Sigma}^{-1}\mv{x}^{(i)}-\frac{1}{2}\log{\left(|\mv{\Sigma}|\right)}-ML_{\rm I}\log{\left(2\pi\right)}-\log{\left(p(\mv{x}^{(i)})\right)}. \label{eq: Sigma2}
\end{align}
Note that $\mv{\Sigma}$ in \eqref{eq:Sigma} is a function of $\mv{\gamma}^{(i-1)}$. In \eqref{eq:grade}, the computation of $\nabla\bar{g}\left(\mv{x}^{(i)}, \mv{\gamma}^{(i-1)}\right) $ is  simple. However, the other term $\exp{\left(\bar{g}\left(\mv{x}^{(i)}, \mv{\gamma}^{(i-1)}\right)\right)}$ in \eqref{eq:grade} often takes either a value which is very close to zero or a pretty large value due to the exponential operator. This problem further leads to the vanishing and exploding gradient problem, which makes the SGD algorithm fragile to be used to solve problem (34). To overcome the above instability issue, we propose to normalize the overall gradient $\nabla g\left(\mv{x}^{(i)}, \mv{\gamma}^{(i-1)}\right)$ at each iteration, i.e., the $\mv{\gamma}^{(i)}$ updated in \eqref{eq: upd} is revised as
\begin{align}
	\mv{\gamma}^{(i)}=\mv{\gamma}^{(i-1)}-\theta_i\frac{\nabla g\left(\mv{x}^{(i)}, \mv{\gamma}^{(i-1)}\right) }{||\nabla g\left(\mv{x}^{(i)}, \mv{\gamma}^{(i-1)}\right)||_2}.\label{eq: upd2}
\end{align}	

In the following, we explain the intuition of the above normalization. Specifically, in \eqref{eq:grade}, denote the gradient direction by
\begin{align}
	\mv{a}^{(i)}=\frac{\nabla g\left(\mv{x}^{(i)}, \mv{\gamma}^{(i-1)}\right)}{||\nabla g\left(\mv{x}^{(i)}, \mv{\gamma}^{(i-1)}\right)||_2}=\frac{\nabla\bar{g}\left(\mv{x}^{(i)}, \mv{\gamma}^{(i-1)}\right)}{||\nabla\bar{g}\left(\mv{x}^{(i)}, \mv{\gamma}^{(i-1)}\right)||_2},
\end{align}
and the gradient length by
\begin{align}
	b^{(i)}=\exp{\left(\bar{g}\left(\mv{x}^{(i)}, \mv{\gamma}^{(i-1)}\right)\right)}||\nabla\bar{g}\left(\mv{x}^{(i)}, \mv{\gamma}^{(i-1)}\right)||_2.
\end{align}
The gradient direction $\mv{a}^{(i)}$ determines the direction along which $\mv{\gamma}^{(i)}$ moves at the $i$-th iteration. The term $b^{(i)}$ determines the step length  with which $\mv{\gamma}^{(i)}$ moves along the corresponding direction at the $i$-th iteration. Note that, here we only study the step length arising from the gradient itself, not the additional parameter $\theta_i$ in  \eqref{eq: upd}. With the step length information $b^{(i)}$, the algorithm might converge quickly. Without the step length information $b^{(i)}$, e.g., the step length is set to be $b^{(i)}=1$, a good performance of the algorithm can still be expected as long as the gradient direction information $\mv{a}^{(i)}$ is provided, and a sufficient number of iterations  is performed. In our case, since the term $\exp{\left(\bar{g}\left(\mv{x}^{(i)}, \mv{\gamma}^{(i-1)}\right)\right)}$ causes significant disturbance of the overall gradient,  we only use the gradient direction information $\mv{a}^{(i)}$ as that in \eqref{eq: upd2}. Note that, compared with the original SGD update in \eqref{eq: upd}, the above normalization may lead to a bias term (i.e., $b^{(i)}$) on the update in \eqref{eq: upd2}. We will show in Section \ref{sec:K=K_ES} that, this bias term does not affect the convergence of the SGD algorithm. We name this algorithm as the Normalized SGD (NSGD) algorithm. The detailed procedures of the NSGD algorithm is shown in Algorithm \ref{alg:ag1}. 

\begin{algorithm}
	\caption{Proposed NSGD Algorithm for Solving Problem \eqref{eq:opt2}}\label{alg:ag1}
	\begin{algorithmic}[1] 
		\State Initialization: $i=0$, $\mv{\gamma}^{(0)}=\mv{0}$, $\eta=1$, $\epsilon>0$.\label{line1}
		\While{$\eta>\epsilon$}
		\State $i\gets i+1$; 
		\State Generate $\mv{x}^{(i)}\in\mathcal{J}\left(\bar{y}_{1}^{Q},\dots,\bar{y}_{2L_{\rm I}M}^{Q}\right)$ in a uniform manner;
		\State Update $\mv{\gamma}^{(i)}$ according to  \eqref{eq: upd2};
		\State Compute $\eta=\frac{||\mv{\gamma}^{(i)}-\mv{\gamma}^{(i-1)}||_1}{N}$;		
		\EndWhile 
		\State Output $\mv{\gamma}^{(i)}$. \label{line14}
	\end{algorithmic}
\end{algorithm} 

The above NSGD algorithm can be viewed as a process that solves a series of optimization problems, each of which is inexactly solved by the traditional gradient descent algorithm with only one gradient descent iteration. That is, instead of solving a problem with objective function being $\mathbb{E}_{\mv{x}}\left(g\left(\mv{x}, \mv{\gamma}\right)\right)$ in \eqref{eq:opt2}, Algorithm \ref{alg:ag1} solves a series of problems with objective function being $g(\mv{x},\mv{\gamma})$.  Specifically, at the $i$-th iteration the problem being solved is
\begin{alignat}{3}
	&\min_{\mv{\gamma}} &\quad &-g\left(\mv{x}^{(i)}, \mv{\gamma}\right) \label{eq:opt2i}\\
	&~~\text{s.t.}                          &      & \mv{\gamma}\ge 0.\nonumber
\end{alignat} 
Performing a normalized gradient descent step on the above problem at the current point $\mv{\gamma}^{(i-1)}$ yields the update in \eqref{eq: upd2}.

\section{Numerical Results of Device Activity Detector}\label{sec:num1}
In this section, we present the device activity detection performance of the proposed NSGD algorithm in terms of missed detection probability (MDP) and false alarm probability (FAP). The MDP is the probability that an active device is detected to be inactive; the FAP is the probability that an inactive device is  detected to be active. The noise power spectrum density is -169 dBm/Hz, the bandwidth is 10 MHz, the transmit power of each device is set to be 23 dBm, and  the large-scale fading component is modeled as  $128.1+37.6\log_{10}(d)$, where $d$ is the distance between the BS and devices in km\cite{3GPP}.  In the simulations, we set $d=1$ km. In addition, all the channel covariance matrices in \eqref{eq:cor} follow the exponential covariance matrix model\cite{loyka2001channel}, i.e., for a covariance matrix $\mv{C}$, the element at the $i$-th row and $j$-th column of $\mv{C}$ is expressed as $[\mv{C}]_{i,j}=c^{i-j}$ if $i\ge j$, and $[\mv{C}]_{i,j}=[\mv{C}]_{j,i}^*$ if $i<j$, with $|c|\le 1$. Moreover, we set $\rho=2$ in \eqref{eq:step2}. Furthermore, we set $\theta_i=i^{-\frac{1}{2}}$ in \eqref{eq: upd2}. To evaluate efficiency of the proposed Algorithm \ref{alg:ag1},  denote $\hat{I}$ the average number of iterations to meet the termination criterion. Last, we have one benchmark to evaluate the detection performance of the proposed NSGD algorithm as follows:

\begin{figure}[h]
	\centering
	\includegraphics[width=8cm]{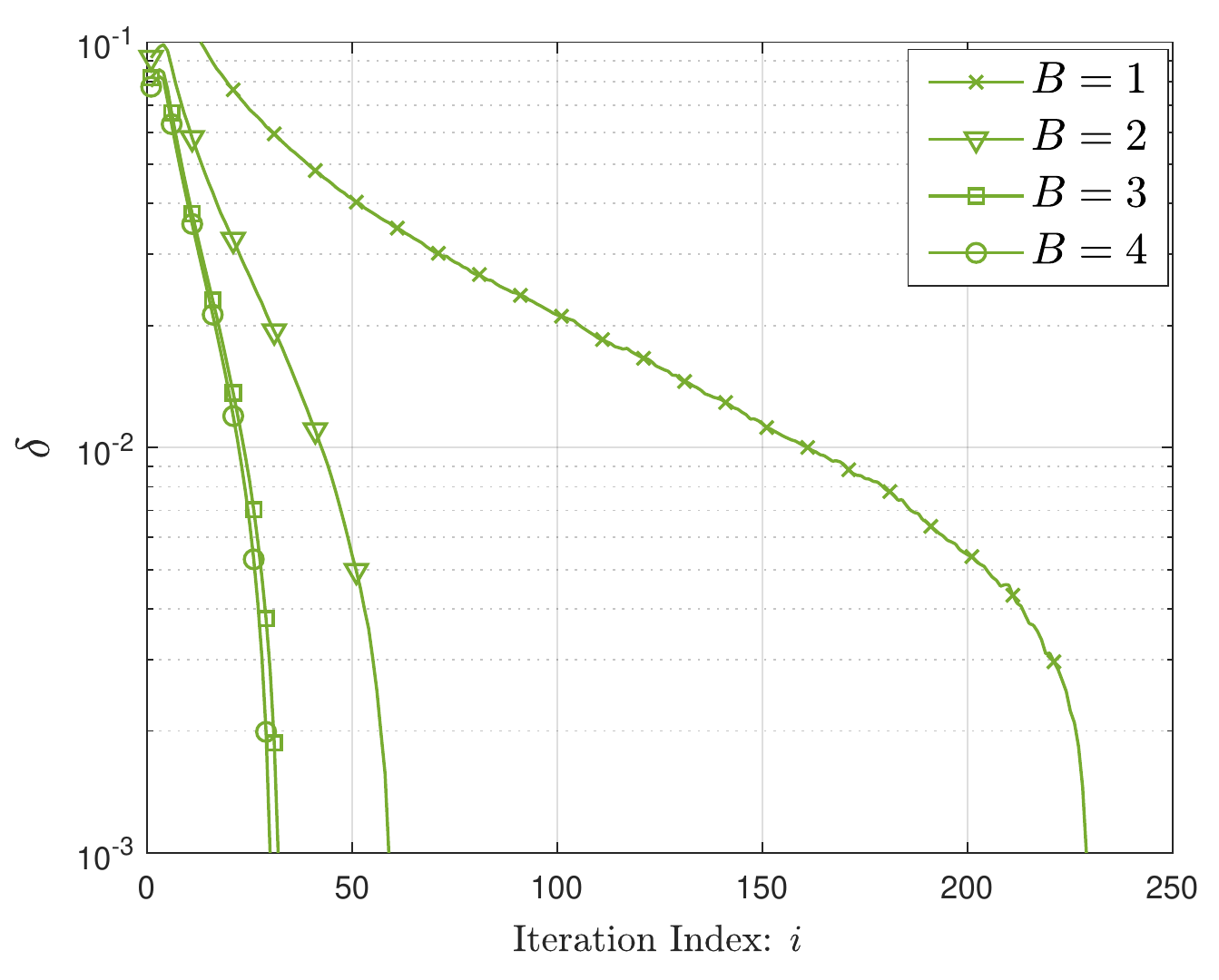}
	\caption{Convergence behavior of proposed Algorithm \ref{alg:ag1}  when $N=1000$, $K=100$, $L_{\rm I}=40$, $M=128$, and $\epsilon=10^{-3}$.}\label{Figconver}
\end{figure}

\begin{itemize}
	\item $B=\infty$. We consider an ADC with infinite-resolution, i.e., the quantization bits $B=\infty$. In this case, the device activity detection problem can be solved by applying the traditional gradient descent algorithm\cite{birgin2000nonmonotone,wang2021efficient}.
\end{itemize}

\subsection{Performance of NSGD with Perfect Knowledge of $K$}\label{sec:K=K_ES}
We first present the performance of NSGD when $\hat{K}=K$. Fig. \ref{Figconver} plots the convergence behavior of the proposed Algorithm \ref{alg:ag1} when $N=1000$, $K=100$, $L_{\rm I}=40$, and $M=128$. In the y-axis in Fig. \ref{Figconver}, denote
\begin{align}
	\delta=\frac{||\mv{\gamma}^{(i)}-\mv{\gamma}^*||_1}{N}
\end{align}
where $\mv{\gamma}^*$ denotes the optimized value of $\mv{\gamma}$ through Algorithm \ref{alg:ag1}. 
We have the following observations from Fig. \ref{Figconver}. First, if we set $\epsilon=10^{-3}$  in \eqref{eq: contol}, the NSGD  indeed converges under different $B$. Second, for the same convergence control parameter $\epsilon$, a larger $B$ can reduce the total number of iterations for the algorithm to meet the termination criterion. For example, the NSGD with $B=3$ converges more quickly than the NSGD with $B=2$. Third, when $B$ is large enough, for the same convergence control parameter $\epsilon$, increasing $B$ can not reduce the number of iterations significantly. For example, the number of iterations required for satisfying the termination criterion are the nearly same when $B=3$ and $B=4$. 


\begin{figure}[h]
	\centering
	\subfigure[Detection performance of NSGD when $N=100$, $K=10$, $L_{\rm I}=13$, and $M=32$.]
	{
		\label{FigSGDPerK1}
		\includegraphics[width=0.45\columnwidth]{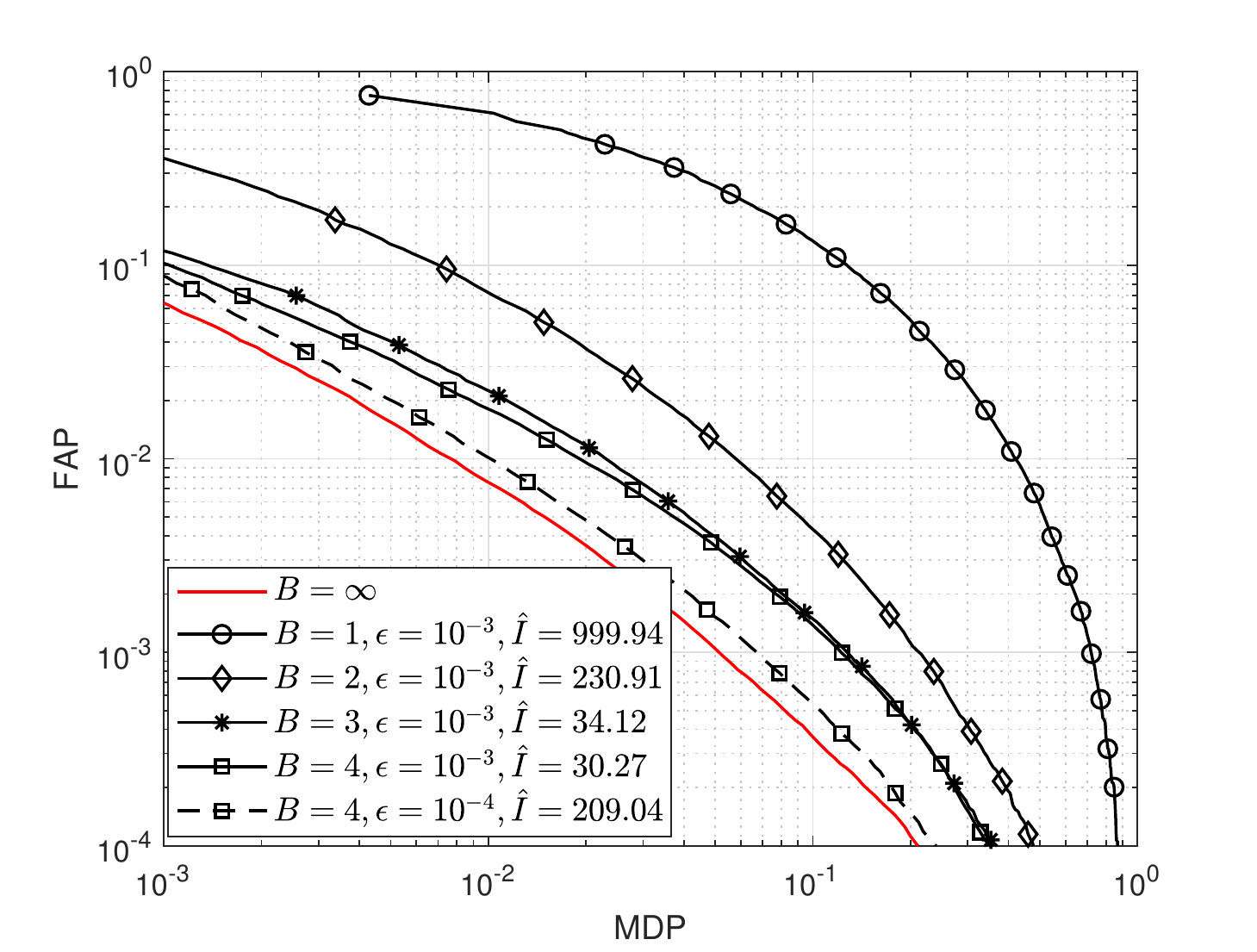}
	}
	\subfigure[Detection performance of NSGD when $N=1000$, $K=100$, $L_{\rm I}=40$, and $M=128$.]
	{
		\label{FigSGDPerK2}
		\includegraphics[width=0.45\columnwidth]{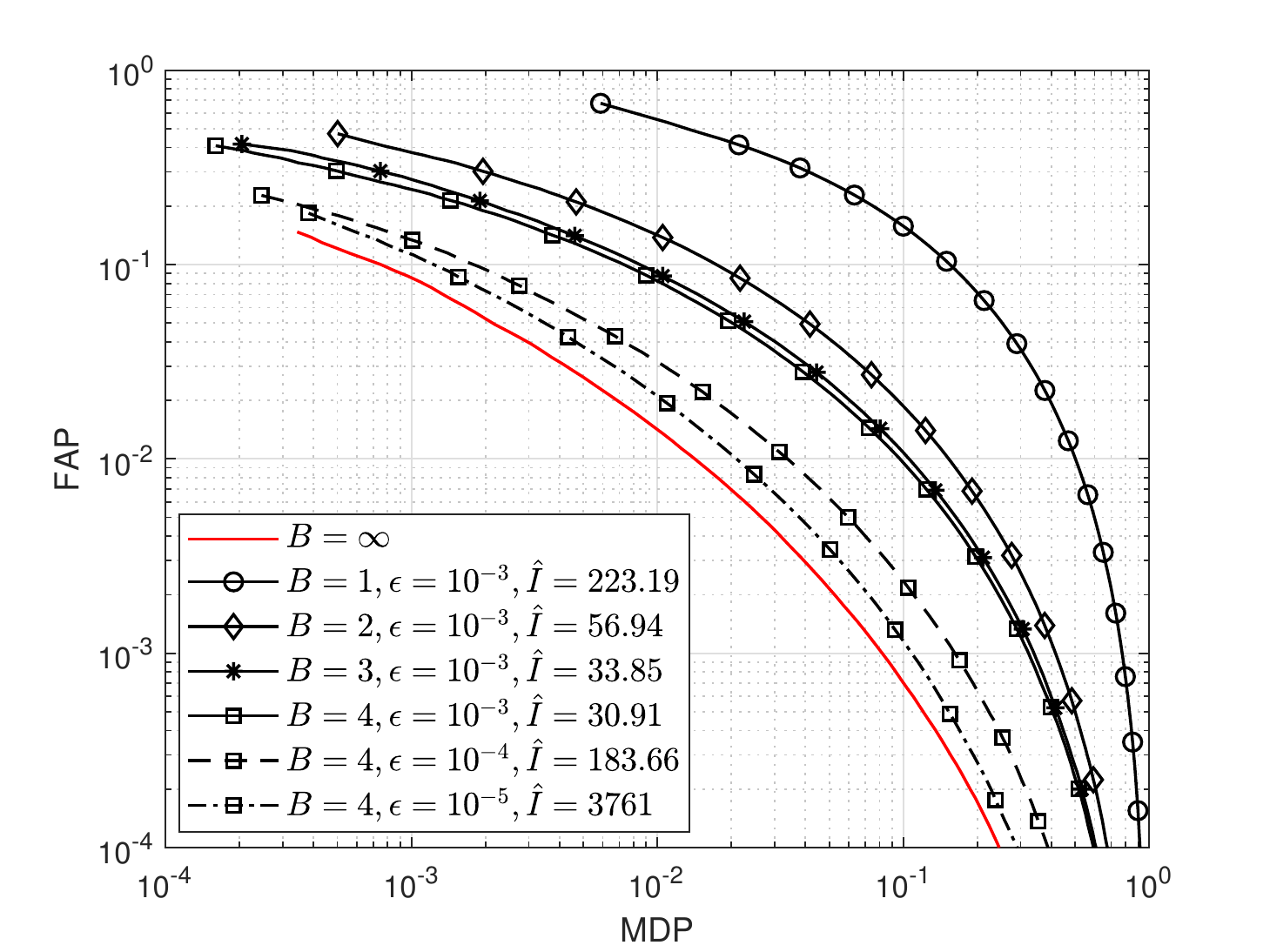}
	}
	\caption{Detection performance of NSGD algorithm  under different setups.}\label{FigSGDPer}
\end{figure}
Next, we present the detection performance of the NSGD algorithm benchmarked against the case with $B=\infty$. First, as shown in Fig. \ref{FigSGDPer}, the performance of NSGD with $B=4$ can approach to the case with infinite ADC resolution. This phenomenon shows that, although the BS is equipped with low-resolution ADCs, in terms of active device detection, we can achieve the performance of the detector with infinite-resolution ADC. Second, the performance of NSGD algorithm increases with the ADC resolution $B$.  Third, improving the solution accuracy (by setting a smaller tolerance $\epsilon$) will require a larger number of iterations. But on the upside it can also possibly improve the detection performance of the proposed NSGD algorithm. Specifically, as shown in Fig. \ref{FigSGDPerK1}, when $\epsilon=10^{-3}$, the performance of NSGD with $B=4$ is roughly the same as that with $B=3$. But if we set $\epsilon=10^{-4}$ or equivalently increase $\hat{I}$ of NSGD from 30.27 to 209.04, we can see that the performance of  NSGD with $B=4$ increases significantly, and even approaches to the performance of the case with infinite ADC resolution. We have similar observations and conclusions on Fig. \ref{FigSGDPerK2}.


\subsection{Performance of NSGD with Imperfect Knowledge of   $K$}\label{sec:Kes}
\begin{figure}[h]
	\centering
	\subfigure[Detection performance of NSGD versus different $\hat{K}$ when $B=4$, $\epsilon =10^{-4}$, $N=1000$, $K=100$, $L_{\rm I}=40$, and $M=128$.]
	{
		\label{FigKes}
		\includegraphics[width=0.45\columnwidth]{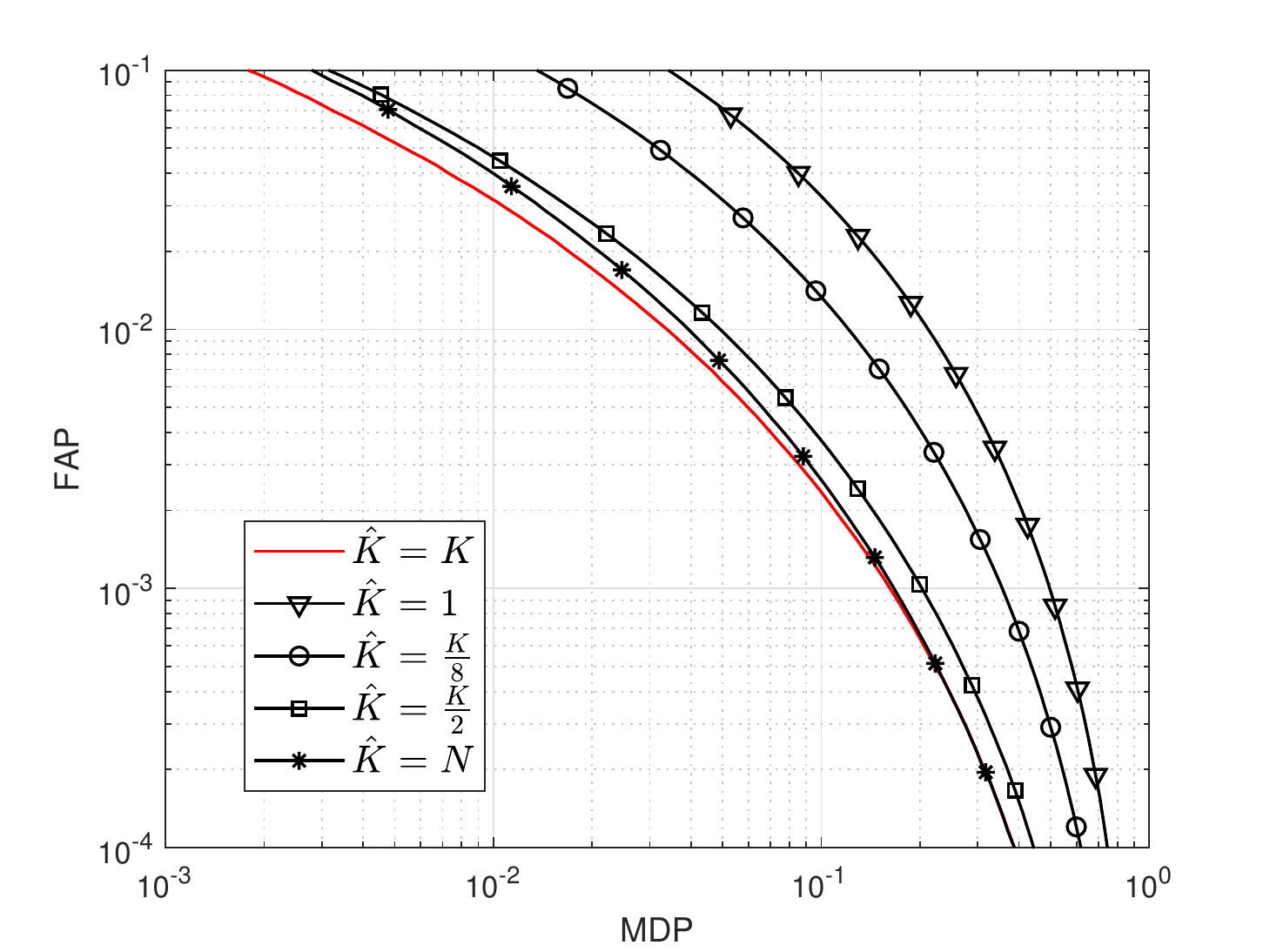}
	}
	\subfigure[Detection performance of NSGD versus different $\hat{K}$ when $B=4$, $\epsilon =10^{-4}$, $N=1000$, $K=10$, $L_{\rm I}=15$, and $M=64$.]
	{
		\label{FigKes2}
		\includegraphics[width=0.45\columnwidth]{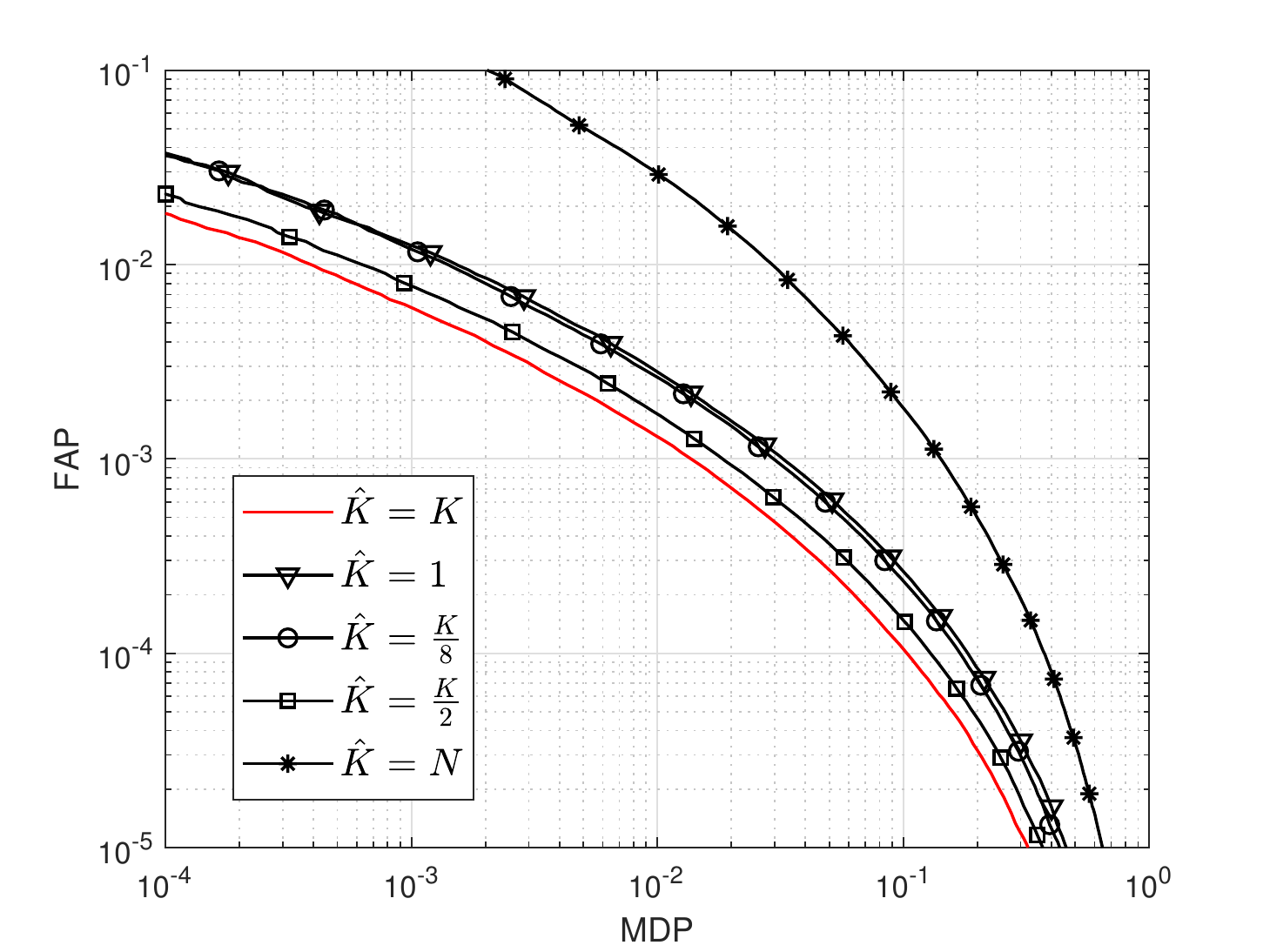}
	}
	\caption{Detection performance of NSGD algorithm versus $\hat{K}$ under different setups.}\label{FigKes3}
\end{figure}

Next, we present the performance of NSGD when only an estimate of $K$, i.e., $\hat{K}$, is known at the BS. The imperfect knowledge of $K$ will affect the estimation of the signal power in  \eqref{eq:lambda21}, and in turn affect the quantization step size design \eqref{eq:step2} as well as the quantization codebook design in  \eqref{eq:codebook}. The numerical results that evaluate the influence of $\hat{K}$ on the NSGD performance are shown in Fig. \ref{FigKes3} when $B=4$. First, Fig. \ref{FigKes3} shows that when $\hat{K}\ne K$, there are detection performance penalties under different $\hat{K}$. For example, there are big performance penalties when $\hat{K}=1$ in Fig. \ref{FigKes}, and when $\hat{K}=N$ in Fig. \ref{FigKes2}. Note that, when assuming $\hat{K}=N$, the detector performs much better in Fig. \ref{FigKes} than that in Fig. \ref{FigKes2}. This is because the designed codebook is the same in Fig. \ref{FigKes} and Fig. \ref{FigKes2} for the  same $\hat{K}$ according to \eqref{eq:lambda21} and \eqref{eq:step2}. Since the gap between $K=100$ and $\hat{K}=N$ is smaller than that between $K=10$ and $\hat{K}=N$, the designed codebook matches the received signals when  $K=100$ in Fig. \ref{FigKes} better than that when $K=10$ in Fig. \ref{FigKes2}. This phenomenon clearly suggests that, the knowledge of $K$ is important for the ADC quantizer design. Otherwise, the mismatch between the ADC quantizer to the received signals will induce big performance penalties. Second, fortunately, the performance of NSGD is still satisfactory when the estimation error of $K$ is not very large. For example, when $\hat{K}=\frac{K}{2}$, there is only a small penalty on the performance of the NSGD in  Fig. \ref{FigKes3}.

In summary, in the case where the knowledge of $K$ and its statistics are unknown at the BS, it is necessary to have an estimation of $K$ at the BS. Fortunately, the estimation does not need to be very accurate, and a fairly good estimation of $K$ is sufficient to achieve satisfactory detection performance.

\section{Number of Active device Estimator in Phase I} \label{sec: es_num}
In this section,  we propose an estimator to estimate the number of active devices $K$ in Phase I of the communication protocol. The received signals to estimate $K$ still need to go through the ADC quantizer, while the design of which still need the information of $K$. Thus, the technical difficulty here is that the design of the ADC quantizer and the estimation of $K$ are closely intertwined and doing one needs the information/execution from the other. This seems like a chicken-and-egg problem. This section aims to propose an efficient algorithm to solve this dilemma.

\subsection{Problem Formulation}
In this section, we study the estimator for $K$ under i.i.d. Rayleigh fading channels. The i.i.d. Rayleigh fading model is widely assumed in the current literature, e.g., \cite{chen2018sparse,liu2018massive,Chenphase,haghighatshoar2018improved,chen2021sparse,chen2019covariance,wang2021efficient,ganesan2020algorithm,accwang,wang2022covariance}. Specifically, following the correlated Rayleigh fading channel model assumed in \eqref{eq:cor}, for the i.i.d. Rayleigh fading channel we have $\mv{C}_n=\mv{I}$ and 
\begin{align}
	\mv{h}_n=\ddot{\mv{h}}_n, ~n=1, 2, \dots,N.\label{eq:cor2}
\end{align}
In Section \ref{sec:es2}, we will show that our estimator designed for i.i.d. Rayleigh fading channel also works well under correlated Rayleigh fading channel.

Since in Phase I we only need to estimate $K$ instead of their identities, the devices can use and send identical preambles, i.e.,  $\mv{s}_1=\dots=\mv{s}_N=\mv{s}\in\mathbb{C}^{L_{\rm N}\times 1}$. 
The received signal in \eqref{eq:rec} is re-expressed as
\begin{align}
	\mv{Y}=\sum_{n=1}^{N}\alpha_n\mv{s}_n\sqrt{\beta}\mv{h}^T_n+\mv{Z}=\mv{s}\sqrt{\beta}\sum_{n=1}^{N}\alpha_n\mv{h}^T_n+\mv{Z}.\label{eq:rec2}
\end{align}
Note that we abuse the notations of the received signals $\mv{Y}$, the preambles $\mv{s}_n$'s, and the other relevant notations in Phase I and Phase II without inducing confusions. In \eqref{eq:rec2}, since the channels follow i.i.d. Rayleigh distribution, we have
\begin{align}
	\sum_{n=1}^{N}\alpha_n\mv{h}^T_n\overset{d}{=}\sqrt{K}\mv{h}^T_1\sim \mathcal{CN}(0,K\mv{I}).
\end{align}
where $\overset{d}{=}$ denotes equal in the distribution. In this case, we have 
\begin{align}
	\mv{Y}\overset{d}{=}\sqrt{K\beta}\mv{s}\mv{h}^T_1+\mv{Z}.\label{eq:rec3}
\end{align}
That is, the received signal $\mv{Y}$ is a function of $K$ in terms of its distribution. The received signal $\mv{Y}$ can be rewritten as a vector form $\bar{\mv{y}}$ like that in \eqref{eq:rec_r} by setting  $\mv{s}_1=\dots=\mv{s}_N=\mv{s}$ therein, then goes through ADC quantizers. The quantized signal is $\bar{\mv{y}}^{Q}$. The estimation problem is thus formulated as
\begin{alignat}{3}
	&\min_{K} &\quad &-p(\bar{\mv{y}}^{Q}
	\mid K,\mv{s}) \label{eq:P4}\\
	&~~\text{s.t.}                          &      & N\ge K\ge 0.\nonumber
\end{alignat}
The objective function is expressed as
\begin{align}
	p(\bar{\mv{y}}^{Q}\mid K,\mv{s})=\int_{\mathcal{I}_{f(\bar{y}_{1}^{Q})}}\int_{\mathcal{I}_{f(\bar{y}_{2}^{Q})}}\dots\int_{\mathcal{I}_{f(\bar{y}_{2L_{\rm N}M}^{Q})}}\prod_{m=1}^{M}p(\hat{\mv{y}}_m\mid K,\mv{s})d\bar{y}_{1}d\bar{y}_{2}\dots d\bar{y}_{2L_{\rm N}M},\label{eq:m2}
\end{align}
where 
\begin{align}
	p(\hat{\mv{y}}_m\mid K,\mv{s})=\frac{1}{(2\pi)^{L_{\rm N}}|\mv{\Sigma_m}|^{\frac{1}{2}}}\exp{\left(-\frac{1}{2}\hat{\mv{y}}_m^T\mv{\Sigma}_m^{-1}\hat{\mv{y}}_m\right)},\label{eq:pdf2}
\end{align}
and $\hat{\mv{y}}_m$ has been similarly defined above \eqref{eq:rec_r}. The covariance matrix of $\hat{\mv{y}}_m$ is expressed as
\begin{align}
	\mv{\Sigma}_m=\mathbb{E}(\mv{\hat{y}}_m\mv{\hat{y}}_m^T)=\frac{1}{2}K\hat{\mv{s}}_1\hat{\mv{s}}_1^T+\frac{1}{2}K\hat{\mv{s}}_{1+N}\hat{\mv{s}}_{1+N}^T+\frac{1}{2}\sigma^2\mv{I}, \label{eq:Sigmam}
\end{align}
where $\hat{\mv{s}}_n$ is the $n$-th column of $\hat{\mv{S}}$ in \eqref{eq:s_hat} when $\mv{s}_1=\dots=\mv{s}_N$.  The estimation problem in \eqref{eq:P4} can be solved by applying the NSGD algorithm detailed in Section \ref{sec:SGD}. We omit the details here.

\subsection{One-Shot Estimator for Number of Active Devices}\label{sec:pea}

Similar to the detector for device activities studied in Section \ref{sec:SGD}, the BS in this section still needs to do ADC quantization once receiving signals from active devices. Then, the BS estimates the number of active devices by solving the estimation problem in  \eqref{eq:P4}. Note that, in order to do the ADC quantization at the BS in Phase I, the BS still needs to have the knowledge of $K$ beforehand.

\begin{figure}[h]
	\centering
	\includegraphics[width=9cm]{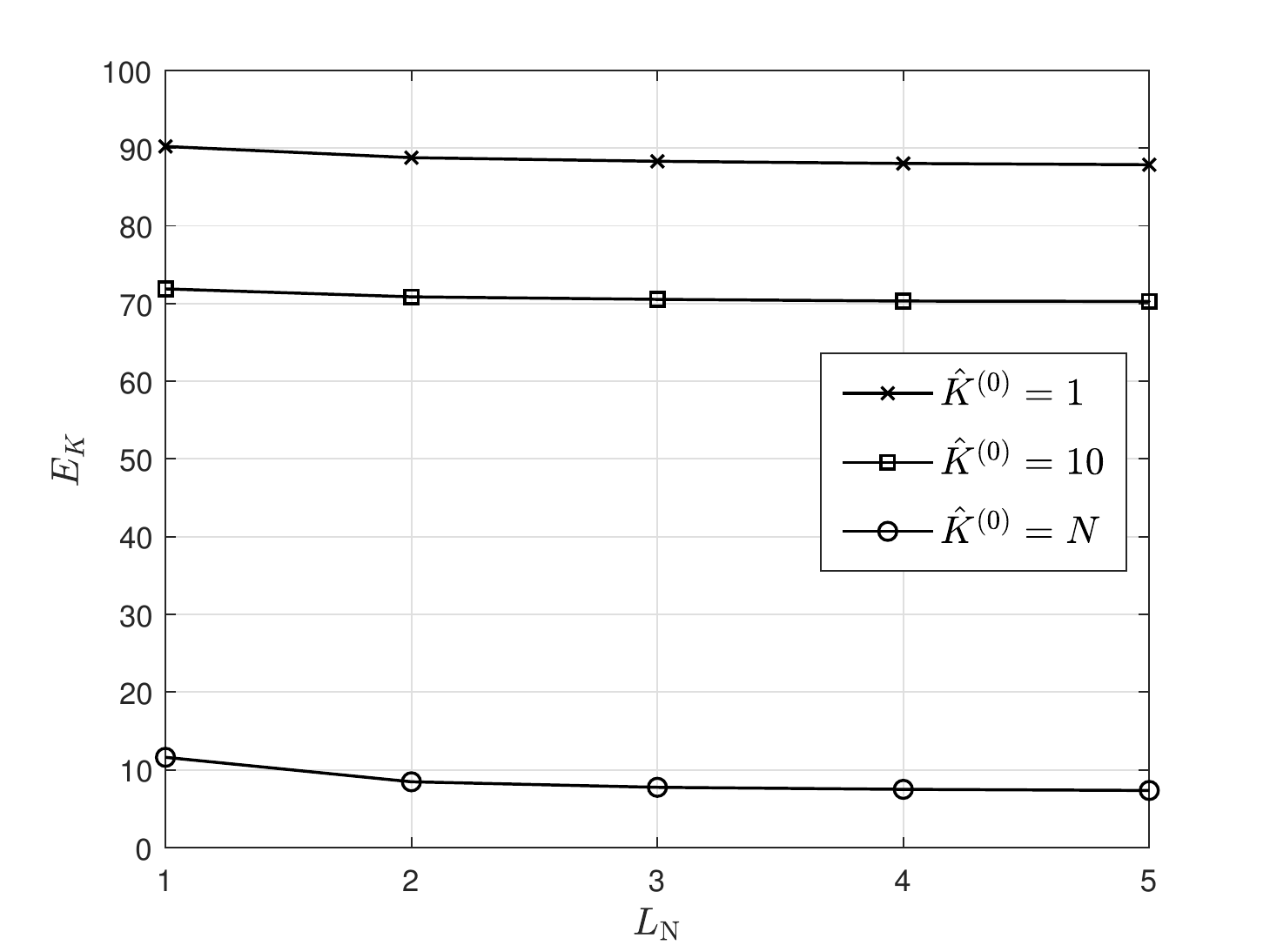}
	\caption{Performance of OEA over different $\hat{K}_{\rm I}$ when $B=4$, $N=1000$, $K=100$, and $M=128$.}\label{Figes} 
\end{figure}

Denote $\hat{K}^{(0)}$ the initial guess that is used to design the ADC quantizer in Phase I. A naive way is that,  we arbitrarily set a value of $\hat{K}^{(0)}$ from 1 to $N$, e.g., $\hat{K}^{(0)}=N$, to  start the estimator in Phase I. Then, BS quantizes signals accordingly and estimates $K$ by solving problem \eqref{eq:P4}. The estimated $K$ is denoted by $\hat{K}$. We name this estimator as \textbf{One-shot Estimator for the number of Active devices (OEA)}. The estimated $K$ through the estimator is delivered to Phase II for detection of the active device identities.

Fig. \ref{Figes} shows the performance of OEA when $B=4$, $N=1000$, $K=100$, and $M=128$. The channel setups are the same as that in  Section \ref{sec:num1}. Denote $E_K$ mean of the estimation error of $K$, and express it as
\begin{align}
	E_K=\mathbb{E}\{|K-\hat{K}|\}.
\end{align}
First, the performance of OEA is sensitive to the initial value $\hat{K}^{(0)}$. Specifically, OEA performs well when $\hat{K}^{(0)}=N$, while it performs poorly when $\hat{K}^{(0)}=1$. The reason for the poor performance of OEA when  $\hat{K}^{(0)}=1$ is that, the estimated sample space in this case is much smaller than the real one in the ADC quantizer. As such, the quantization error would be quite large. Second, the estimation error decreases slowly with $L_{\rm N}$. This shows that, in order to reduce the estimation error, increasing $L_{\rm N}$ is not a good idea. Much worse, if too much time is spent on the estimation of $K$, the time for the identity detection would be quite limited.

\subsection{Progressive Estimator for Number of Active Devices}\label{sec:pea2}
We observe from Fig. \ref{Figes} that when $\hat{K}^{(0)}=1$, the estimation error of OEA is
\begin{align}
	E_K=\mathbb{E}\{|100-\hat{K}|\}\approx90.\label{eq:err1}
\end{align}
If we treat the initial guess $\hat{K}^{(0)}=1$ as the estimated value as well, then the estimation error would be
\begin{align}
		E_K=\mathbb{E}\{|100-\hat{K}^{(0)}|\}=99,\label{eq:err}
\end{align}
which is larger than the estimation error of OEA in \eqref{eq:err1}. That is, although the estimation error of OEA is pretty large, OEA can help reduce the estimation error arising from the initial value as in \eqref{eq:err}. This is also true if the initial value is chosen to be $10$. 

Building upon this observation, we propose a \textbf{Progressive Estimator for the number of Active devices (PEA)}. PEA turns the one-shot estimation in OEA into a series of progressive estimations. Specifically, at  the $i$-th estimation, each device only uses one symbol $s_{i}\in\mathbb{C}$ to estimate $K$, where $s_{i}$ denotes the $i$-th element in $\mv{s}$. In this case, for PEA, the number of estimations equals the number of preamble length $L_{\rm N}$. Denote the estimated $K$ at the $i$-th estimation by $\hat{K}^{(i)}$, $i=1,\dots, L_{\rm N}$.  Then, the procedures of the PEA  are summarized in Algorithm \ref{alg:ag2}.

\begin{algorithm}
	\caption{Proposed PEA Algorithm}\label{alg:ag2}
	\begin{algorithmic}[1] 
		\State Initialization: $L_{\rm N}$, $\hat{K}^{(0)}$.
		\For{$i=1,\dots, L_{\rm N}$}
		\State According to $\hat{K}^{(i-1)}$, the BS designs the ADC quantizer based on \eqref{eq:codebook}; 
		\State Each device transmits symbol $s_i$ to estimate $K$; 
		\State The BS quantizes the received signals $\bar{\mv{y}}$ according to the designed ADC quantizer;
		\State The BS estimates $K$ by solving the problem \eqref{eq:P4}, the output of which is $\hat{K}^{(i)}$.		
		\EndFor 
		\State Output $\hat{K}=\hat{K}^{(L_{\rm N})}$ that is delivered to Phase II for the detection of device activities. 
	\end{algorithmic}
\end{algorithm}

\begin{figure}[t]
	\centering
	\subfigure[Estimated $K$ over different initial values $\hat{K}^{(0)}$.]
	{
		\label{ERR1}
		\includegraphics[width=0.45\columnwidth]{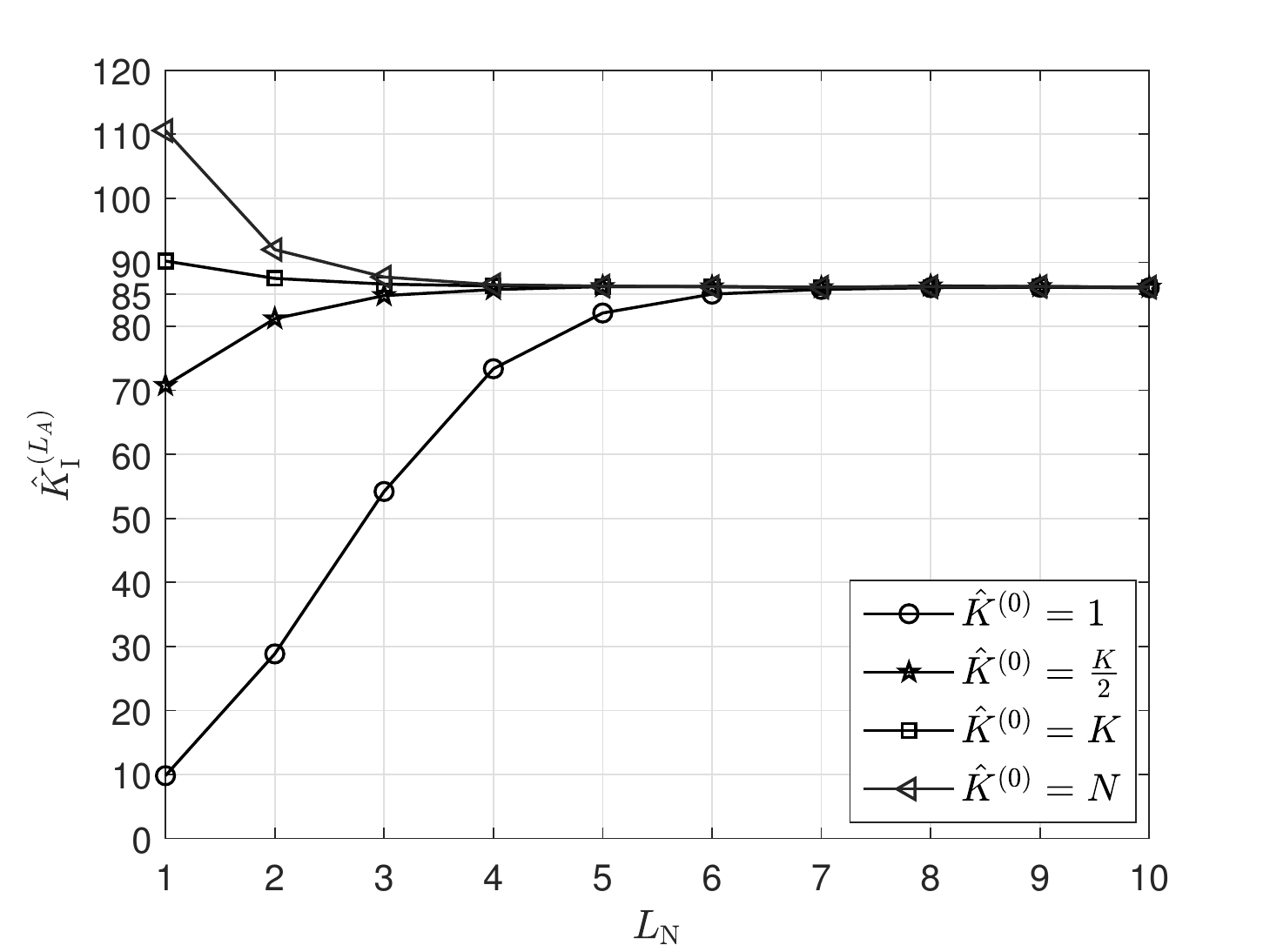}
	}
	\subfigure[Estimation error of $K$ over different initial values $\hat{K}^{(0)}$.]
	{
		\label{ERR2}
		\includegraphics[width=0.45\columnwidth]{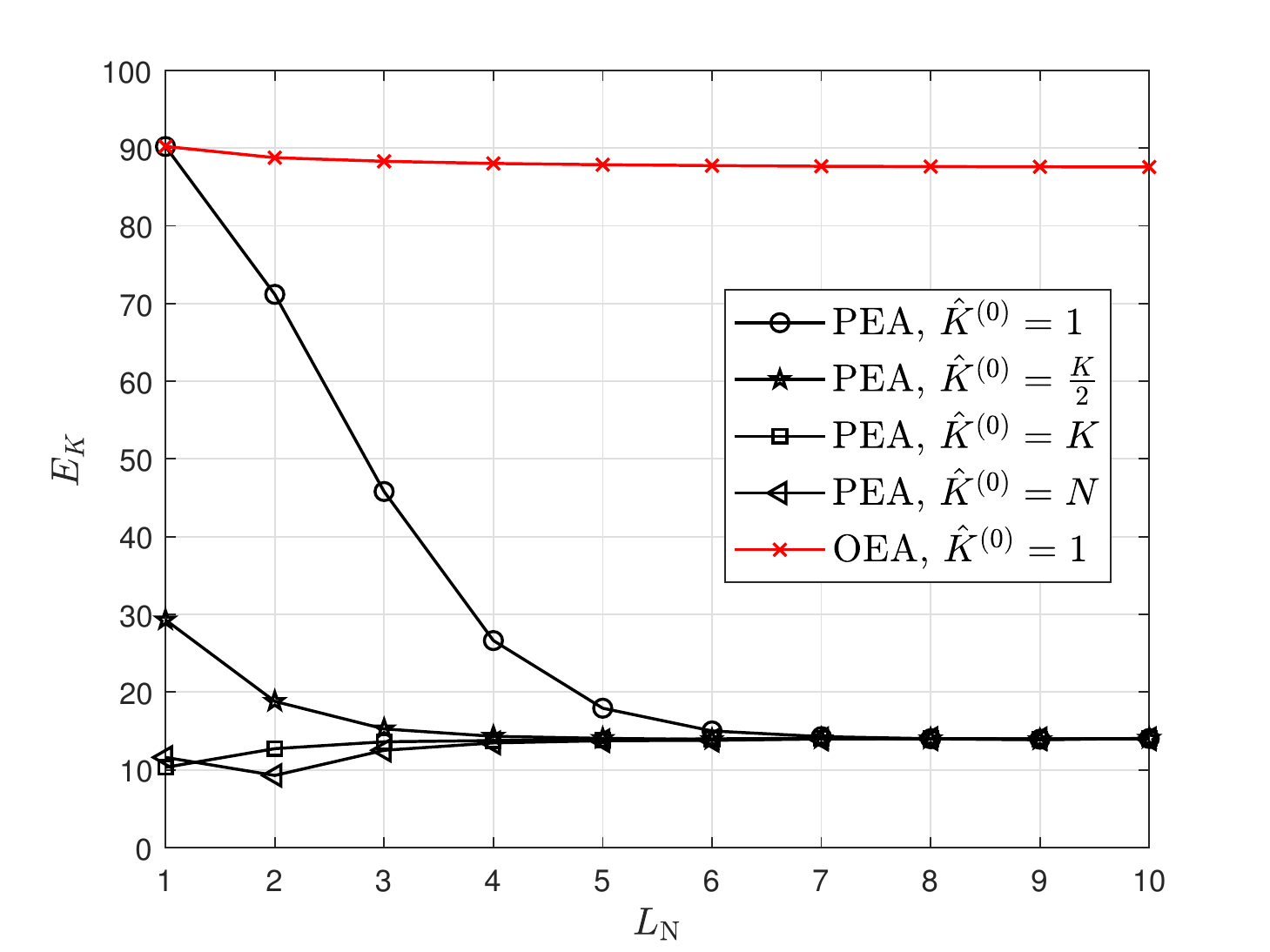}
	}
	\caption{Estimation performance of PEA when $B=4$, $N=1000$, $K=100$, and $M=128$.}\label{ERR}
\end{figure}

\subsection{Performance of PEA under I.I.D. Rayleigh Fading Channel}\label{sec:es}
In this section, we present the numerical results of the proposed estimator PEA for the number of active devices, and the overall performance of the communication protocol in Fig. \ref{Figpro}  when $B=4$, $N=1000$, $K=100$, and $M=128$. The channel setups are the same as that in  Section \ref{sec:num1}. 

Fig. \ref{ERR1} shows the estimated $K$ over different initial values $\hat{K}^{(0)}$. First, when $\hat{K}^{(0)}=1$ and $L_{\rm N}\le 6$, the output value of PEA is always closer to the real $K$ than the input value, i.e.,
\begin{align}
	\mathbb{E}\{|K-\hat{K}^{(i)}|\}<\mathbb{E}\{|K-\hat{K}^{(i-1)}|\}, ~i=1,\dots, L_{\rm N}.
\end{align}
Also, as shown in Fig. \ref{ERR2}, the estimation error of $K$ decreases as $L_{\rm N}$ increases when $\hat{K}^{(0)}=1$. Benchmarked against OEA (which is sensitive to the initial value) when the initial value is set to be 1, PEA has significant performance improvement. Second, as shown in Fig. \ref{ERR1}, for different $\hat{K}^{(0)}$, the estimated $K$ can converge to $\hat{K}^{(L_{\rm N})}=85$ as the increase of $L_{\rm N}$. This phenomenon suggests that our PEA is quite robust to the choice of initial value  $\hat{K}^{(0)}$. The finally converged value $\hat{K}^{(L_{\rm N})}=85$ may be mainly determined by the number of antennas $M$ and the quantization bits $B$. We leave it as a future work to analyze the convergence and  the influence factors on the converged value.

\begin{figure}[ht]
	\centering
	\includegraphics[width=8cm]{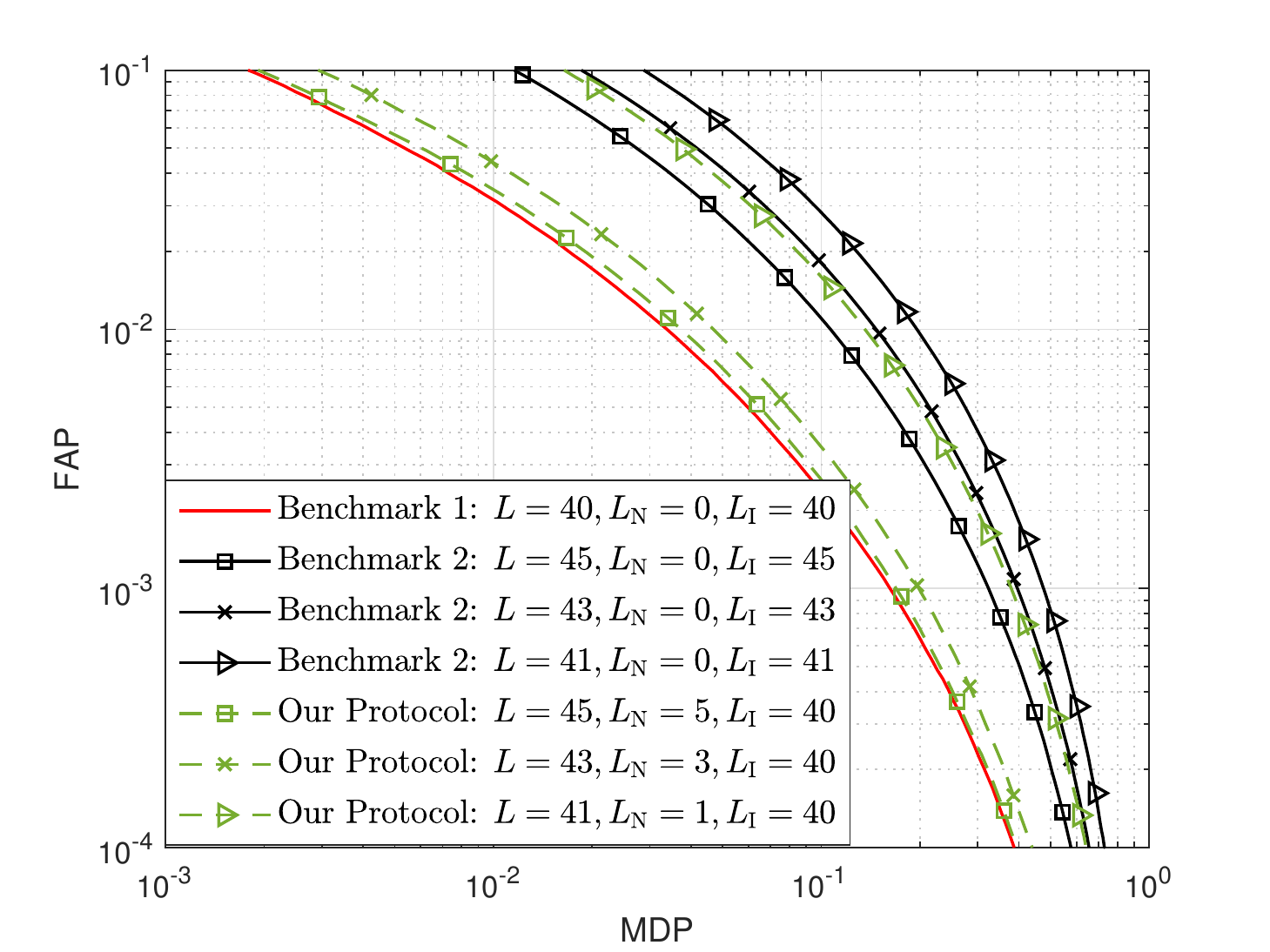}
	\caption{Detection performance of the communication protocol in Fig. \ref{Figpro}.}\label{Fig1}
\end{figure}

In Fig. \ref{Fig1}, we show the performance of the proposed communication protocol in Fig. \ref{Figpro} that consists of an estimator for the number of active devices and a detector for device identity. To evaluate its performance, we have the following to benchmarks:
\begin{enumerate}
	\item \textbf{Benchmark 1}: It only consists of a detector for device identity, and has perfect knowledge of $K$, i.e.,  $\hat{K}=K$.
	\item \textbf{Benchmark 2}: It only consists of a detector for device identity, and simply assumes  $\hat{K}=1$.
\end{enumerate}
As shown in Fig. \ref{Fig1}, our proposed communication protocol performs quite well. First, for the same overall preamble length $L$, compared with Benchmark 2, our protocol has significant improvement on the detection performance since we allocate some preambles to estimate $K$ while Benchmark 2 simply assumes $\hat{K}=1$.  That is, although we use less number of preambles to detect user identities in Phase II,  we can still achieve the overall performance improvement compared with Benchmark 2, since the estimation of $K$ is good enough. Second, compared with Benchmark 1 which has perfect knowledge of $K$, by using additional 5 preambles to estimate $K$, our protocol performs nearly the same as Benchmark 1. This clearly shows that the proposed PEA performs well and the estimated $K$ is good enough for the detector in Phase II. 

\subsection{Performance of PEA under Correlated Rayleigh Fading Channel}\label{sec:es2}

\begin{figure}[ht]
	\centering
	\includegraphics[width=9cm]{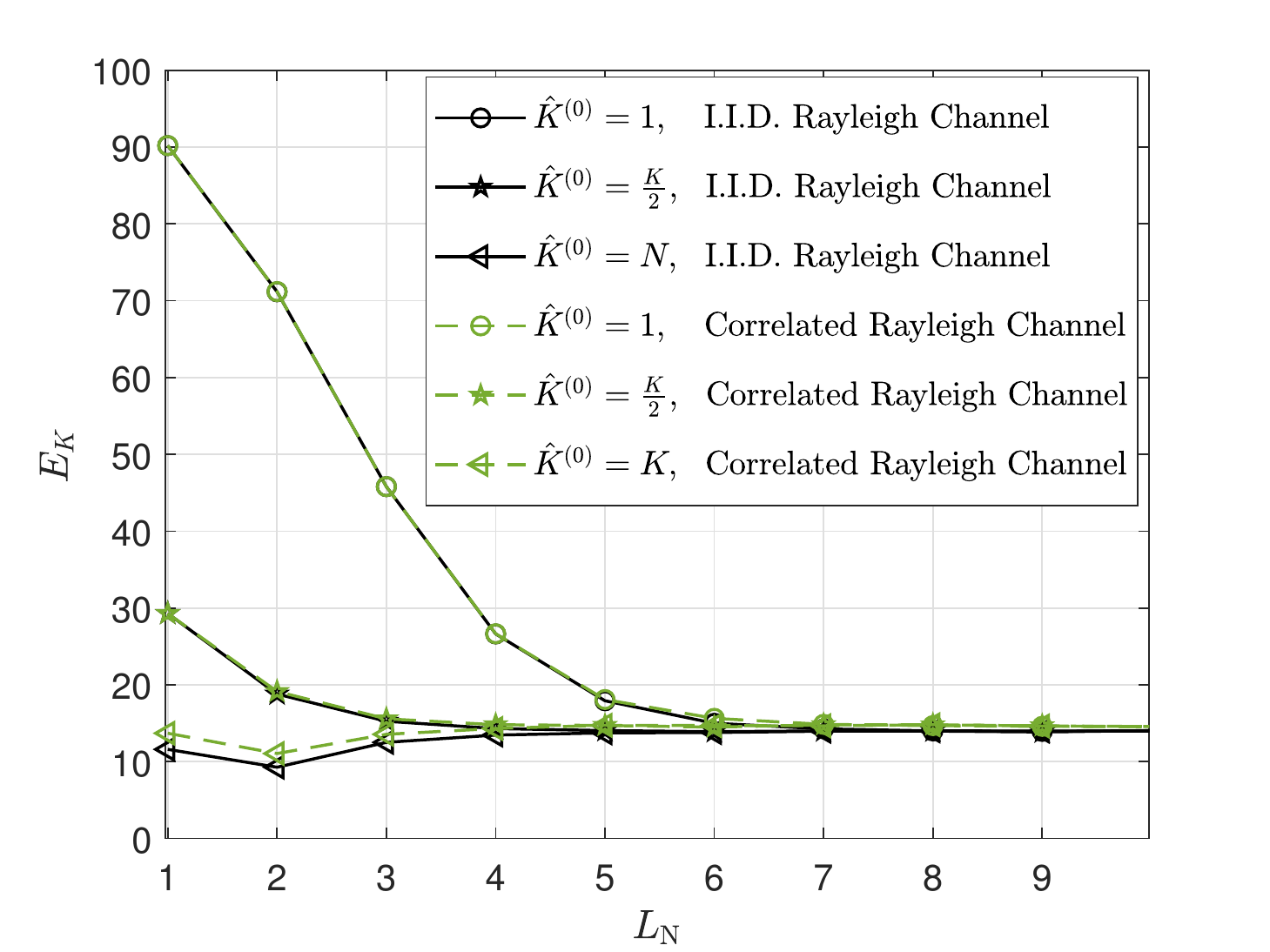}
	\caption{Estimation performance of PEA when $B=4$, $N=1000$, $K=100$, and $M=128$ under correlated Rayleigh fading channel.}\label{Fig12}
\end{figure}
In Fig. \ref{Fig12}, under correlated Rayleigh fading channels, we show the performance of PEA that has been designed for i.i.d. Rayleigh fading channels in Section \ref{sec:pea}. The channel setups and the channel covariance model are the same as that in Section \ref{sec:num1}. Fig. \ref{Fig12} shows that the performance of PEA under correlated Rayleigh fading channels is roughly the same as that under i.i.d. Rayleigh fading channels, which thus shows that our designed estimator for the number of active devices is quite robust to the channel distributions. The possible reason is that the proposed estimator PEA does not have high  estimation accuracy. In this case, the prior information of the channel distribution is not important for the estimator. 

\section{Conclusion}\label{sec:col}
We investigated a device activity detection problem in mMTC, where the BS is equipped with a massive MIMO and low-resolution ADCs. This setup causes additional challenges in the activity detection problem. First, the BS should have the knowledge of the number of active devices $K$ in order to do better ADC quantization. Second, it is not clear if the covariance-based approach can still achieve good performance. To solve the above problems, we proposed a communication protocol that consists of an estimator for $K$ and a detector for active device identities. For the  estimator for $K$, we proposed a progressive estimator PEA which iteratively performs the estimation of  $K$ and the design of the quantizer. For the detector for active device identities, we proposed a custom-designed NSGD algorithm to solve the problem whose objective involves high-dimensional integration. Numerical results demonstrated the effectiveness of the proposed communication protocol.

\begin{appendix}
\subsection{Proof of Theorem \ref{theorem0}}\label{appendix0}
First, from \eqref{eq:rec}, the received signal $\mv{y}_m$ at the $m$-th antenna is
\begin{align}
	\mv{y}_m=\mv{S}\mv{\gamma}^{\frac{1}{2}}\mv{h}^{\rm col}_m+\mv{z}_m, ~m=1, 2, \dots, M,\label{eq:rec_m}
\end{align}
where $\mv{h}^{\rm col}_m$ denotes the $m$-th column of $\mv{H}$ in \eqref{eq:rec}. Second, to express \eqref{eq:rec_m} into a real form, denote
$\hat{\mv{y}}_m=\left[\Re{(\mv{y}^T_m)}, \Im{(\mv{y}^T_m)}\right]^T\in\mathbb{R}^{2L_{\rm I}\times 1}$,  $\hat{\mv{S}}=\begin{bmatrix}
	\Re{\left(\mv{S}\right)} & -\Im{\left(\mv{S}\right)}\\
	\Im{\left(\mv{S}\right)} & \Re{\left(\mv{S}\right)}
\end{bmatrix}\in\mathbb{R}^{2L_{\rm I}\times 2N}$, $\hat{\mv{\gamma}}={\rm diag}\left\{\mv{\gamma}, \mv{\gamma}\right\}$,  $\hat{\mv{h}}_m=\left[\Re{(\mv{h}^{\rm col}_m)^T}, \Im{(\mv{h}^{\rm col}_m)^T}\right]^T\in\mathbb{R}^{2N\times 1}$, and $\hat{\mv{z}}_m=\left[\Re{(\mv{z}^T_m)}, \Im{(\mv{z}^T_m)}\right]^T\in\mathbb{R}^{2L_{\rm I}\times 1}$. In this case, \eqref{eq:rec_m} can be rewritten as
\begin{align}
	\hat{\mv{y}}_m=\hat{\mv{S}}\hat{\mv{\gamma}}^{\frac{1}{2}}\hat{\mv{h}}_m+\hat{\mv{z}}_m, ~m=1, 2, \dots, M,\label{eq:rec_mr}
\end{align}
where $\hat{\mv{z}}_m\sim \mathcal{N}(0,\frac{1}{2}\sigma^2\mv{I})$, $\forall m$. 
Last, denote $\bar{\mv{y}}=\left[\hat{\mv{y}}_1^T,\dots, \hat{\mv{y}}_M^T\right]^T\in\mathbb{R}^{2L_{\rm I}M\times 1}$,  $\bar{\mv{S}}={\rm diag}\{\underbrace{\hat{\mv{S}},\dots,\hat{\mv{S}}}_M\}\in\mathbb{R}^{2L_{\rm I}M\times 2NM}$ which consists of $M$ copies of $\hat{\mv{S}}$ in its main-diagonal blocks, $\bar{\mv{\gamma}}={\rm diag}\{\underbrace{\hat{\mv{\gamma}},\dots,\hat{\mv{\gamma}}}_M\}\in\mathbb{R}^{2NM\times 2NM}$, $\bar{\mv{h}}=\left[\hat{\mv{h}}_1^T,\dots, \hat{\mv{h}}_M^T\right]^T\in\mathbb{R}^{2NM\times 1}$, and $\bar{\mv{z}}=\left[\hat{\mv{z}}_1^T,\dots, \hat{\mv{z}}_M^T\right]^T\in\mathbb{R}^{2L_{\rm I}M\times 1}$. Then, the received signal $\mv{Y}$ in \eqref{eq:rec} is rewritten as
\begin{align}
	\bar{\mv{y}}=\bar{\mv{S}}\bar{\mv{\gamma}}^{\frac{1}{2}}\bar{\mv{h}}+\bar{\mv{z}}.\label{eq:rec_r2}
\end{align}	
\subsection{Derivations of $\mv{\Sigma}$ and $\mv{C}$ in \eqref{eq:Sigma}}\label{appendix1}

We first derive the  $\mv{\Sigma}$ in \eqref{eq:Sigma}. Since the first equality in \eqref{eq:Sigma} is straightforward, we mainly focus on the proof of the second equality. To this end, we first take a close look at the covariance matrix $\mv{C}=\mathbb{E}(\mv{\bar{h}}\mv{\bar{h}}^T)$ in \eqref{eq:Sigma}. Specifically, from Section \ref{sec: sys_A}, we know
\begin{align}
	\bar{\mv{h}}=\left[\hat{\mv{h}}_1^T, \dots, \hat{\mv{h}}_M^T\right]^T=\left[\Re{(\mv{h}^{\rm col}_1)^T}, \Im{(\mv{h}^{\rm col}_1)^T},  \dots, \Re{(\mv{h}^{\rm col}_M)^T}, \Im{(\mv{h}^{\rm col}_M)^T}\right]^T, \label{eq:h}
\end{align}
where $\mv{h}^{\rm col}_m$ denotes the $m$-th column of $\mv{H}$ in \eqref{eq:rec}. It follows from \eqref{eq:h} that
\begin{align}
	\bar{\mv{h}}_{2m-1}=\Re{(\mv{h}^{\rm col}_m)}, ~ \bar{\mv{h}}_{2m}=\Im{(\mv{h}^{\rm col}_m)},~m=1,\dots, M.  \label{eq:de}
\end{align} 
The covariance matrix of $\bar{\mv{h}}$ is further expressed as
\begin{align}
	\mv{C}=\mathbb{E}(\mv{\bar{h}}\mv{\bar{h}}^T)=\begin{bmatrix}
		\mv{C}_{1,1} & \mv{C}_{1,2}    & \cdots  & \mv{C}_{1,2M}\\
		\vdots   & \vdots    & \ddots  & \vdots\\
		\mv{C}_{2M,1}   & \mv{C}_{2M,2}    & \cdots  & \mv{C}_{2M,2M}
	\end{bmatrix}, \label{eq:C}
\end{align}
where
\begin{align}
	\mv{C}_{m,\bar{m}}&=\mathbb{E}(\bar{\mv{h}}_m\bar{\mv{h}}_{\bar{m}}^T)\label{eq:C0}\\
	&=\mathbb{E}\left(\left[\bar{h}_{1,m},\dots,\bar{h}_{N,m}\right]^T\left[\bar{h}_{1,\bar{m}},\dots, \bar{h}_{N,\bar{m}}\right]\right), ~ m, \bar{m}=1,\dots, 2M, \label{eq:C1}
\end{align}
where $\bar{h}_{n,m}$ denotes the $n$-th element in $\bar{\mv{h}}_{m}$. Now let us take a close look at $\mv{C}_{m,\bar{m}}$. Recall that
$\bar{h}_{n,m}$ is the channel of device $n$, and $\bar{h}_{\bar{n},\bar{m}}$ is the channel of device $\bar{n}$. Since the channels among different devices are independent, we have  $\mathbb{E}\left(\bar{h}_{n,m}\bar{h}_{\bar{n},\bar{m}}\right)=0$,  $\forall n\neq \bar n$. As a result,  $\mv{C}_{m,\bar{m}}$'s in \eqref{eq:C0} are diagonal matrices. Based on this fact,  in \eqref{eq:Sigma} we have 
\begin{align}
	\mv{\gamma}^{\frac{1}{2}}\mv{C}_{m,\bar{m}}\mv{\gamma}^{\frac{1}{2}}=\mv{C}_{m,\bar{m}}\mv{\gamma}, ~~\forall m, \bar{m}.\label{eq: gam}
\end{align}
Combining \eqref{eq: gam} with the definition of $\mv{\Sigma}$ (i.e., the first equality in \eqref{eq:Sigma}) gives
\begin{align}
	\mv{\Sigma}=&\bar{\mv{S}}\begin{bmatrix}
		\mv{C}_{1,1}\mv{\gamma} & \mv{C}_{1,2}\mv{\gamma}   & \cdots  & \mv{C}_{1,2M}\mv{\gamma}\\
		\vdots   & \vdots    & \ddots  & \vdots\\
		\mv{C}_{2M,1}\mv{\gamma}   & \mv{C}_{2M,2}\mv{\gamma}   & \cdots  & \mv{C}_{2M,2M}\mv{\gamma}
	\end{bmatrix}\bar{\mv{S}}^T+\frac{1}{2}\sigma^2\mv{I}\label{eq:Sigma2}\\
	&=\bar{\mv{S}}\mv{C}\bar{\mv{\gamma}}\bar{\mv{S}}^T+\frac{1}{2}\sigma^2\mv{I}. \label{eq:Sigma3}
\end{align}
This completes the derivation of $\mv{\Sigma}$ in \eqref{eq:Sigma}.

Next, we derive the expression of $\mv{C}$ in \eqref{eq:C}. Basically, we want to build connections between the elements in $\mv{C}=\mathbb{E}(\mv{\bar{h}}\mv{\bar{h}}^T)$ and the elements in  $\mv{C}_n=\mathbb{E}(\mv{h}_n\mv{h}_n^T)$'s in \eqref{eq:cor}. To this end, we first show the relationship between $\mv{h}_n$ and $\bar{\mv{h}}$.  We do this through the following two steps:
\begin{enumerate}
	\item The relationship between $\mv{h}_n$ and $\mv{h}_m^{\rm col}$ in $\bar{\mv{h}}$, $n=1,\dots, N$, $m=1,\dots,M$. Recall that $\mv{h}_m^{\rm col}$ denotes the $m$-th column of $\mv{H}=[\mv{h}_1, \mv{h}_2,\dots, \mv{h}_N]^T$. Then we have	 
	\begin{align}
		\mv{h}_n^T=\left[h_{n,1}^{\rm col},\dots,h_{n,M}^{\rm col}\right],
	\end{align}
    where $h_{n,m}^{\rm col}$ is the $n$-th element in $\mv{h}_m^{\rm col}$. Moreover, rewriting $\mv{h}_n$ in a real form gives
    \begin{align}
    		\left[\Re{\left(\mv{h}_n^T\right)}, \Im{\left(\mv{h}^T_n\right)}\right]^T
    		=\left[\Re{\left(h_{n,1}^{\rm col}\right)},\dots,\Re{\left(h_{n,M}^{\rm col}\right)}, \Im{\left(h_{n,1}^{\rm col}\right)},\dots,\Im{\left(h_{n,M}^{\rm col}\right)}\right]. \label{eq:rel3}	
    \end{align}
    \item The relationship between $\mv{h}_m^{\rm col}$ and $\bar{\mv{h}}_m$ in $\bar{\mv{h}}$, $m=1,\dots, M$.  Specifically, according to the definition in  \eqref{eq:de}, we have
    \begin{align}
    	\Re{\left(h_{n,m}^{\rm col}\right)}=\bar{h}_{n,2m-1}, ~ \Im{\left(h_{n,m}^{\rm col}\right)}=\bar{h}_{n,2m}, ~m=1,\dots,M. \label{eq: 68}
    \end{align}

Then, substituting \eqref{eq: 68} into \eqref{eq:rel3}, the relationship $\mv{h}_n$ and $\bar{\mv{h}}$ can be  expressed as
\begin{align}
	\left[\Re{\left(\mv{h}_n^T\right)}, \Im{\left(\mv{h}^T_n\right)}\right]^T
	=\left[\bar{h}_{n,1}, \bar{h}_{n,3}, \dots, \bar{h}_{n,2M-1}, \bar{h}_{n,2}, \bar{h}_{n,4}, \dots, \bar{h}_{n,2M}\right]^T, ~n=1,\dots,N. \label{eq:cn}
\end{align}
\end{enumerate}

Second, we show the relationship between the elements in $\mv{C}=\mathbb{E}(\mv{\bar{h}}\mv{\bar{h}}^T)$ and the elements in  $\mv{C}_n=\mathbb{E}(\mv{h}_n\mv{h}_n^T)$'s  through \eqref{eq:cn}. Note that, from \eqref{eq:cor} we have
\begin{align}
	\left[\Re{\left(\mv{h}_n^T\right)}, \Im{\left(\mv{h}^T_n\right)}\right]^T
	=\bar{\mv{C}}_n^{\frac{1}{2}}\left[\Re{\left(\ddot{\mv{h}}_n^T\right)}, \Im{\left(\ddot{\mv{h}}^T_n\right)}\right]^T, \label{eq:re}
\end{align}
where
\begin{align}
	\bar{\mv{C}}_n^{\frac{1}{2}}=\begin{bmatrix}
		\Re{\left(\mv{C}_n^{\frac{1}{2}}\right)} & -\Im{\left(\mv{C}_n^{\frac{1}{2}}\right)}\\
		\Im{\left(\mv{C}_n^{\frac{1}{2}}\right)} & \Re{\left(\mv{C}_n^{\frac{1}{2}}\right)}
	\end{bmatrix},\label{eq:cnbar}
\end{align}
$\Re{\left(\ddot{\mv{h}}_n\right)}\sim \mathcal{N}(0,\frac{1}{2}\sigma^2\mv{I})$, and $\Im{\left(\ddot{\mv{h}}_n\right)}\sim \mathcal{N}(0,\frac{1}{2}\sigma^2\mv{I})$. Combining \eqref{eq:cn}, \eqref{eq:re}, and  \eqref{eq:cnbar} together, we have
\begin{align}
    &\mathbb{E}\left(\left[\bar{h}_{n,1}, \dots, \bar{h}_{n,2M-1}, \bar{h}_{n,2},  \dots, \bar{h}_{n,2M}\right]^T\left[\bar{h}_{n,1}, \dots, \bar{h}_{n,2M-1}, \bar{h}_{n,2},  \dots, \bar{h}_{n,2M}\right]\right)\nonumber\\
	&=\mathbb{E}\left(\left[\Re{\left(\mv{h}_n^T\right)}, \Im{\left(\mv{h}^T_n\right)}\right]^T\left[\Re{\left(\mv{h}_n^T\right)}, \Im{\left(\mv{h}^T_n\right)}\right]\right)\nonumber\\
	&=\bar{\mv{C}}_n^{\frac{1}{2}}\frac{1}{2}\mv{I}\bar{\mv{C}}_n^{\frac{1}{2}}=\frac{1}{2}\bar{\mv{C}}_n, ~~\forall n. \label{eq:corre3}
\end{align}
Denote $c^{n,n}_{m,\bar{m}}$ the $n$-th diagonal element of $\mv{C}_{m,\bar{m}}$ in \eqref{eq:C}. From \eqref{eq:corre3},  for $k, \bar{k} =1,\dots, M$, the $n$-th diagonal element of the diagonal matrix $\mv{C}_{m,\bar{m}}$ is expressed as
\begin{align}
	c^{n,n}_{m,\bar{m}}=\mathbb{E}\left(\bar{h}_{n,m}\bar{h}_{n,\bar{m}}\right)=\left\{\begin{array} {llll} \frac{1}{2}\bar{c}_n^{k,\bar{k}}, ~~~~~~~{\rm if} ~ m=2k-1, \bar{m}=2\bar{k}-1, \\ \frac{1}{2}\bar{c}_n^{k,\bar{k}+M}, ~~~~{\rm if} ~ m=2k-1, \bar{m}=2\bar{k}, \\\frac{1}{2}\bar{c}_n^{k+M,\bar{k}}, ~~~~{\rm if} ~ m=2k, \bar{m}=2\bar{k}-1,\\\frac{1}{2}\bar{c}_n^{k+M,\bar{k}+M}, ~{\rm if} ~ m=2k, \bar{m}=2\bar{k}, \end{array}\right. \label{eq:rela}
\end{align}
Though \eqref{eq:rela}, we can construct $\mv{C}$ from $\mv{C}_n$'s.

\subsection{Proof of Theorem \ref{theorem1}}\label{appendix2}
Let $\lambda^{\ell, \ell}$ denote the power of $\bar y_{\ell}$,~$\ell=1,2,\dots, 2L_{\rm I}M$, where  $\bar{y}_{\ell}$ denotes the $\ell$-th element of the received signal  $\bar{\mv{y}}$ in \eqref{eq:rec_r}. Since  $\mv{\Sigma}$  is the covariance matrix of $\bar{\mv{y}}$, we compute the diagonal elements of $\mv{\Sigma}$, i.e., $\lambda^{\ell,\ell}$, $\forall l$. In \eqref{eq:Sigma}, denote
\begin{align}
	\mv{D}=\bar{\mv{S}}\mv{C}\bar{\mv{\gamma}}\bar{\mv{S}}^T\in\mathbb{R}^{2L_{\rm I}M\times 2L_{\rm I}M}, \label{eq:D2}
\end{align}
and $d^{\ell,\ell}$ the $\ell$-th diagonal element of $\mv{D}$. According to \eqref{eq:Sigma}, we have
\begin{align}
	\lambda^{\ell,\ell}=d^{\ell,\ell}+\frac{1}{2}\sigma^2,~~ \ell=1,\dots, 2L_{\rm I}M. \label{lambda}
\end{align}
In the following, we compute $d^{\ell,\ell}$. Substituting $\bar{\mv{S}}={\rm diag}\{\underbrace{\hat{\mv{S}},\dots,\hat{\mv{S}}}_M\}$, $\mv{C}$ in \eqref{eq:C}, and $\bar{\mv{\gamma}}={\rm diag}\{\underbrace{\hat{\mv{\gamma}},\dots,\hat{\mv{\gamma}}}_M\}$  into $\mv{D}$ in \eqref{eq:D2}, we have
\begin{align}
	\mv{D}=\bar{\mv{S}}\mv{C}\bar{\mv{\gamma}}\bar{\mv{S}}^T=\begin{bmatrix}
		\mv{D}_{1,1} & \mv{D}_{1,2}   & \cdots  & \mv{D}_{1,M}\\
		\vdots   & \vdots    & \ddots  & \vdots\\
		\mv{D}_{M,1}   & \mv{D}_{M,2}   & \cdots  & \mv{D}_{M,M}
	\end{bmatrix} \label{eq:D}
\end{align}
where 
\begin{align}
	\mv{D}_{m,\bar{m}}=\hat{\mv{S}}\begin{bmatrix}
		\mv{C}_{2m-1, 2\bar{m}-1}\mv{\gamma} & \mv{C}_{2m-1, 2\bar{m}}\mv{\gamma}\\
		\mv{C}_{2m, 2\bar{m}-1}\mv{\gamma} & \mv{C}_{2m, 2\bar{m}}\mv{\gamma}
	\end{bmatrix}\hat{\mv{S}}^T,~~ m, \bar{m}=1,\dots, M, \label{eq:Dm}
\end{align}
and the expression of $\hat{\mv{S}}$ is shown below \eqref{eq:rec_m}. Since we want to compute the diagonal elements of $\mv{D}$ in \eqref{eq:D}, we only care about the diagonal elements in $\mv{D}_{m,m}$ in \eqref{eq:Dm}, $m=1,\dots, M$.

First, we show that within \eqref{eq:Dm}, we have
\begin{align}
	\mv{C}_{2m-1,2m}=\mv{C}_{2m,2m-1}=\mv{0}, m=1,\dots, M.
\end{align}
We first show $\mv{C}_{2m-1,2m}=\mv{0}$, $m=1,\dots, M$. Since $\mv{C}_{2m-1,2m}$ is a diagonal matrix, it suffices to show that all diagonal elements of $\mv{C}_{2m-1,2m}$ are zero. According to \eqref{eq:rela}, the $n$-th diagonal element in $\mv{C}_{2m-1,2m}$ is expressed as
\begin{align}
	c_{2m-1,2m}^{n,n}=\frac{1}{2}\bar{c}^{m,m+M}_n, n=1,\dots, N. \label{eq: 46}
\end{align}
In the following, we compute $\bar{c}^{m,m+M}_n$'s in $\bar{\mv{C}}_n$. The $\bar{\mv{C}}_n$ in \eqref{eq:cnbar} is expressed as
\begin{align}
	\bar{\mv{C}}_n=\left(\bar{\mv{C}}_n\right)^{\frac{1}{2}}\left(\bar{\mv{C}}_n\right)^{\frac{1}{2}}&=\begin{bmatrix}
		\mv{A}_n & -\mv{B}_n\\
		\mv{B}_n & \mv{A}_n
	\end{bmatrix}\begin{bmatrix}
	\mv{A}_n & -\mv{B}_n\\
	\mv{B}_n & \mv{A}_n
\end{bmatrix}\nonumber\\
&=\begin{bmatrix}
\mv{A}_n\mv{A}_n-\mv{B}_n\mv{B}_n & -\mv{A}_n\mv{B}_n-\mv{B}_n\mv{A}_n\\
\mv{A}_n\mv{B}_n+\mv{B}_n\mv{A}_n & \mv{A}_n\mv{A}_n-\mv{B}_n\mv{B}_n
\end{bmatrix}, \label{eq: R0}
\end{align}
where $\mv{A}_n=\Re{\left(\mv{C}_n^{\frac{1}{2}}\right)}$ and $\mv{B}_n=\Im{\left(\mv{C}_n^{\frac{1}{2}}\right)}$. In addition, since
\begin{align}
	\mv{C}_n&=\Re{\left(\mv{C}_n\right)}+i\Im{\left(\mv{C}_n\right)}=\mv{C}_n^{\frac{1}{2}}\mv{C}_n^{\frac{1}{2}}=\left(\mv{A}_n+i\mv{B}_n\right)\left(\mv{A}_n+i\mv{B}_n\right)\nonumber\\&=\mv{A}_n\mv{A}_n-\mv{B}_n\mv{B}_n+i\left(\mv{A}_n\mv{B}_n+\mv{B}_n\mv{A}_n\right), ~\forall n,
\end{align}
 we have 
\begin{align}
	\Re{\left(\mv{C}_n\right)}=\mv{A}_n\mv{A}_n-\mv{B}_n\mv{B}_n, ~\Im{\left(\mv{C}_n\right)}=\mv{A}_n\mv{B}_n+\mv{B}_n\mv{A}_n. \label{eq: R1}
\end{align}
Substituting \eqref{eq: R1} into 
\eqref{eq: R0}, we have
\begin{align}
	\bar{\mv{C}}_n=\begin{bmatrix}
		\Re{\left(\mv{C}_n\right)} & -\Im{\left(\mv{C}_n\right)}\\
		\Im{\left(\mv{C}_n\right)} & \Re{\left(\mv{C}_n\right)}
	\end{bmatrix}, n=1,\dots, N.\label{eq: R3}
\end{align}
Denote  $c_n^{m,m}$ the $m$-th diagonal element of $\mv{C}_n$. Since $c_n^{m,m}=1$, $m=1,\dots,M$, in \eqref{eq: R3} we have
\begin{align}
	\bar{c}^{m,m+M}_n=-\Im{\left(c_n^{m,m}\right)}=0,
\end{align}
which, together with \eqref{eq: 46}, further implies
\begin{align}
	c_{2m-1,2m}^{n,n}=\frac{1}{2}\bar{c}^{m,m+M}_n=0, n=1,\dots, N. \label{eq: 47}
\end{align}
Thus, $\mv{C}_{2m-1,2m}=\mv{0}$, $m=1,\dots, M$. Similarly, we can show that $\mv{C}_{2m,2m-1}=\mv{0}$, $m=1,\dots, M$.

By combining the facts that $\mv{C}_{2m-1,2m}$'s and $\mv{C}_{2m,2m-1}$'s  are zero matrices,  we can further express $\mv{D}_{m,m}$ in \eqref{eq:Dm} as
\begin{align}
	\mv{D}_{m,m}&=\sum_{n=1}^Nc^{2m-1, 2m-1}_{n,n}\gamma_{n}\hat{\mv{s}}_n\hat{\mv{s}}_n^T+c^{2m, 2m}_{n,n}\gamma_{n}\hat{\mv{s}}_{n+N}\hat{\mv{s}}_{n+N}^T\label{eq:54}\\&=\frac{1}{2}\sum_{n=1}^N\gamma_{n}\hat{\mv{s}}_n\hat{\mv{s}}_n^T+\gamma_{n}\hat{\mv{s}}_{n+N}\hat{\mv{s}}_{n+N}^T, \forall m, \label{eq:55}
\end{align}
where $\gamma_{n}$ denotes the $n$-th element of $\mv{\gamma}$ in \eqref{eq:rec}, and $\hat{\mv{s}}_n$ is the $n$-th column of $\hat{\mv{S}}$. In addition, \eqref{eq:54} to \eqref{eq:55} follows from the fact that, according to  \eqref{eq: 46}, we have $c^{2m-1, 2m-1}_{n,n}=c^{2m, 2m}_{n,n}=\frac{1}{2}$, $m=1\dots, M$, and $N=1,\dots, N$. Denote $d_{m,m}^{\ell,\ell}$ the $\ell$-th diagonal element in $\mv{D}_{m,m}$. We have 
\begin{align}
	d^{\ell,\ell}_{m,m}=\frac{1}{2}\sum_{n=1}^N\gamma_{n}\hat{s}_{\ell, n}\hat{s}_{\ell, n}+\frac{1}{2}\sum_{n=1}^N\gamma_{n}\hat{s}_{\ell, n+N}\hat{s}_{\ell, n+N}, ~m=1,\dots, M, ~\ell=1, \dots, 2L_{\rm I}, \label{eq:pw}
\end{align}
where $\hat{s}_{\ell, n}$ denotes the element in the $\ell$-th row and $n$-th column of $\hat{\mv{S}}$. Within \eqref{eq:pw}, for the first term in the right-hand side, we have
\begin{align}
	\frac{1}{2}\sum_{n=1}^N\gamma_{n}\hat{s}_{\ell, n}\hat{s}_{\ell, n}\overset{(a)}{=}\frac{1}{4}\sum_{n=1}^N\gamma_{n}=\frac{1}{4}\sum_{n\in\mathcal{K}}\beta=\frac{K}{4}\beta,  \label{eq:term1}
\end{align}
where the equality $(a)$ follows from the fact that $|\hat{s}_{\ell, n}|^2=\frac{1}{2}$, and the set $\mathcal{K}=\{n: \alpha_n=1, \forall n\}$ denotes the set containing the indices of  $K$ active devices.  Similarly, for the second term in the right-hand side of \eqref{eq:pw},  we have
\begin{align}
	\frac{1}{2}\sum_{n=1}^N\gamma_{n}\hat{s}_{\ell, n+N}\hat{s}_{\ell, n+N}=\frac{1}{4}\sum_{n\in\mathcal{K}}\beta=\frac{K}{4}\beta. \label{eq:term2}
\end{align}
Substituting \eqref{eq:term1} and \eqref{eq:term2} into \eqref{eq:pw}, we have
\begin{align}
	d^{\ell,\ell}_{m,m}=\frac{K}{2}\beta,  ~~m=1,\dots, M, ~\ell=1, \dots, 2L_{\rm I}.\label{eq:d}
\end{align}
Thus, we have 
\begin{align}
	d^{\ell,\ell}=\frac{K}{2}\beta,  ~~\ell=1, \dots, 2L_{\rm I}M.\label{eq:d2}
\end{align}
Substituting \eqref{eq:d2} into \eqref{lambda},  we have 
\begin{align}
	\lambda^{\ell,\ell}=d^{\ell,\ell}+\frac{1}{2}\sigma^2=\frac{K}{2}\beta+\frac{1}{2}\sigma^2,~~ \ell=1,\dots, 2L_{\rm I}M. \label{lambda21}
\end{align}

\end{appendix}

\bibliographystyle{IEEEtran}
\bibliography{CIC}

\end{document}